\documentclass[bibyear]{aa} 

\usepackage{subcaption}
\usepackage{graphicx}
\usepackage{txfonts}

\usepackage[usenames,table,dvipsnames]{xcolor}
\definecolor{blue}{RGB}{66, 153, 233}
\definecolor{red}{RGB}{255, 0, 0}
\definecolor{purple}{RGB}{255, 0, 255}
\usepackage{placeins}

\usepackage{natbib,twoopt}

\usepackage[breaklinks=true]{hyperref} 
\bibpunct{(}{)}{;}{a}{}{,}             
\makeatletter
  \newcommandtwoopt{\citeads}[3][][]{\href{http://adsabs.harvard.edu/abs/#3}%
    {\def\hyper@linkstart##1##2{}%
     \let\hyper@linkend\@empty\citealp[#1][#2]{#3}}}
  \newcommandtwoopt{\citepads}[3][][]{\href{http://adsabs.harvard.edu/abs/#3}%
    {\def\hyper@linkstart##1##2{}%
     \let\hyper@linkend\@empty\citep[#1][#2]{#3}}}
  \newcommandtwoopt{\citetads}[3][][]{\href{http://adsabs.harvard.edu/abs/#3}%
    {\def\hyper@linkstart##1##2{}%
     \let\hyper@linkend\@empty\citet[#1][#2]{#3}}}
  \newcommandtwoopt{\citeyearads}[3][][]%
    {\href{http://adsabs.harvard.edu/abs/#3}
    {\def\hyper@linkstart##1##2{}%
     \let\hyper@linkend\@empty\citeyear[#1][#2]{#3}}}
\makeatother

\let\oldthebibliography\thebibliography
\renewcommand{\thebibliography}[1]{%
  \oldthebibliography{#1}
  \let\oldbibitem\bibitem
  \let\oldtextsc\textsc
  \def\oldbbland{et}
  \newcounter{authorcount}
  \def\bibitem[##1]##2{%
    \let\textsc\oldtextsc
    \let\bbland\oldbbland
    \oldbibitem[##1]{##2}%
    \let\textsc\mytextsc%
    \let\bbland\mybbland
    \setcounter{authorcount}{0}
  }
  \def\mybbland{\setcounter{authorcount}{0}\oldbbland}
  \def\dropetal##1.{ \bbletal}
  \def\mytextsc##1{%
    \oldtextsc{##1}%
    \stepcounter{authorcount}%
    \ifnum\value{authorcount}=3\relax%
      \expandafter\dropetal%
    \fi%
  }%
}

\begin{document} 

\title{A Unified Multi-Wavelength Data Analysis Workflow with \texttt{gammapy}}
\subtitle{Constraining the Broadband Emission of FSRQ OP 313}

\author{M. Nievas Rosillo
          \inst{1,2}
          \and
        F. Acero
          \inst{3,4}
          \and
        J. Otero-Santos
          \inst{5,6}
          \and
        M. Vazquez Acosta
          \inst{1,2}
          \and \\
        R. Terrier
          \inst{7} 
          \and
        D. Morcuende
          \inst{5}
          \and
        A. Arbet-Engels
          \inst{8}
        }

\institute{Instituto de Astrofísica de Canarias (IAC), C/ Vía Láctea s/n, 38205 La Laguna, Tenerife, Spain \\
              \email{mnievas@iac.es}
         \and
            Universidad de La Laguna (ULL), Avda. Astrofísico Francisco Sánchez s/n, 38206 La Laguna, Tenerife, Spain
         \and
            FSLAC IRL 2009, CNRS/IAC, La Laguna, Tenerife, Spain
         \and 
            AIM, CEA, CNRS, Université Paris-Saclay, Université de Paris, F-91191 Gif-sur-Yvette, France
         \and
            Instituto de Astrofísica de Andalucía (IAA-CSIC), Glorieta de la Astronomía s/n, 18008 Granada, Spain
         \and
            Istituto Nazionale di Fisica Nucleare, Sezione di Padova, 35131 Padova, Italy
         \and
            APC, Université de Paris, CNRS, CEA, Observatoire de Paris, 10 rue Alice Domon et Léonie Duquet, 75013 Paris, France
         \and
            Max Planck Institute for Physics (MPP), Boltzmannstr. 8, 85748 Garching/Munich, Germany
         }

   \date{Received September 23, 2024; accepted XX XX, 202X}

 
  \abstract
   {The Flat Spectrum Radio Quasar OP~313 entered an enhanced activity phase in November 2023 and has undergone multiple flares since then, which have motivated the organization of several large multi-wavelength campaigns, including two deep observations from the hard X-ray telescope NuSTAR. The broadband emission from OP~313 during these two observations is investigated under a new unified analysis framework, with data spanning from optical to $\gamma$ rays.} 
   {Traditional methods for analyzing blazar emission often rely on proprietary software tailored to specific instruments, making it challenging to integrate and interpret data from multi-wavelength campaigns comprehensively. This study demonstrates the feasibility of utilizing \texttt{gammapy}, an open-source Python package, and common data formats originally developed for $\gamma$-ray instrumentation, to perform a consistent multi-instrument analysis. This enables a forward folding approach that fully incorporates source observations, detector responses, and various instrumental and astrophysical backgrounds. The methodology is applied as an example to recent data collected from the distant quasar OP~313.}
   {We present a comprehensive data reconstruction and analysis for instruments including the Liverpool Telescope's IO:O detector, {\it Swift}-UVOT, {\it Swift}-XRT, NuSTAR, and \textit{Fermi}-LAT. The resulting spectral analysis is validated against the native tools for each instrument. Additionally, a multi-wavelength phenomenological model of the source emission, encompassing optical to $\gamma$-ray bands, is developed, incorporating absorption components across different energy regimes.}
   {We introduce and validate a new unified framework for multi-wavelength forward folding data analysis based on \texttt{gammapy} and open data formats, demonstrating its application to spectral data from the quasar OP~313. This approach provides a more statistically correct treatment of the data than fitting a collection of flux points extracted from the different instruments. This study is the first to use a common event data format and analysis tool covering 11 orders of magnitude in energy, from approximately 1 eV to 100 GeV. The high-level event data, instrument response functions, and models are provided in a \texttt{gammapy}-compatible format, ensuring accessibility and reproducibility of scientific results. A brief discussion on the origin of OP~313's broadband emission is also included.}
  {}

   \keywords{blazars: general --
             galaxies: active --
             radiation mechanisms: non-thermal --
             methods: data analysis --
             gamma rays: galaxies
             }

   \maketitle

\section{Introduction}

Active Galactic Nuclei (AGNs), in particular those with large-scale jets of ultra-relativistic particles, are the predominant extragalactic sources of $\gamma$-ray photons. Their non-thermal emission extends over the entire electromagnetic spectrum, from radio to $\gamma$~rays; yet the driving physical mechanism that explains the radiation in each band is different. At low energies, from radio to optical (in some cases up to X-rays), the emission is often attributed to synchrotron radiation from relativistic electrons \citep{koenigl1981}. At higher energies, the situation is less clear and diverse mechanisms are assumed to explain the emission, from inverse Compton of the same electrons with the synchrotron radiation \citep[synchrotron-self-Compton or SSC, see][]{maraschi1992} or with external thermal photons from the AGN structure \citep[external Compton or EC, e.g.][]{dermer1993}, to various hadronic processes: proton-synchrotron, proton-proton \citep[see e.g.][]{aharonian2002}.

Flat Spectrum Radio Quasars (FSRQs) are a subclass of AGNs characterized by jets closely aligned with the 
line of sight and strong thermal radiation components from the AGN accretion disk in the optical-UV band. This radiation is partially reprocessed into thermal emission by the dusty torus in the infrared and into Doppler-broadened emission lines in the broad line region (BLR). These thermal radiation fields are significant targets for Compton (up)scattering, making FSRQs often very luminous in the $\gamma$-ray band, especially during flares. This process is a major cooling mechanism for the accelerated electrons, as the electrons lose energy by scattering photons to higher energies. The high luminosity and close jet alignment makes FSRQs detectable at greater distances compared to BL Lac objects, which lack strong thermal components.

However, detecting very-high-energy (VHE; E $>$ 100 GeV) $\gamma$-ray emission from AGNs --- especially for those found at high redshifts ---
is challenging 
due to attenuation by the extragalactic background light \citep[EBL, e.g.][]{dominguez2011, saldana-lopez2021}, 
which absorbs $\gamma$ rays from distant sources and is an irreducible background that encodes important information about the star formation history in the Universe. The induced absorption produces an energy-dependent imprint on the blazar spectrum, reducing significantly the observable $\gamma$-ray flux in the VHE band. At the same time, it provides an opportunity to constrain the density of the EBL indirectly, provided that we can infer the intrinsic spectrum of the source. For sources at redshift $z\sim 1$ and sub-TeV photons, the most relevant part of the spectrum of the EBL is the so-called Cosmic Optical Background component.

OP~313 is a FSRQ at a redshift of $z=0.997$ \citep{schneider2010}, which has experienced several flaring states over the past 15 years, as evidenced by the continuous monitoring with the \textit{Fermi}-LAT $\gamma$-ray space telescope. In November 2023, OP~313 entered a multi-month high-state, with daily energy flux often exceeding $\mathrm{F(>\,100\, MeV) \gtrsim 10^{-10}\, erg\, cm^{-2}\,s^{-1}}$ in the {\it Fermi} light curve repository \citep{2023ApJS..265...31A}, representing the brightest flare for this source since \textit{Fermi}'s launch in 2008, more than two orders of magnitude above the quiescent state flux. Around February 29, 2024 (MJD60369), the source reached a record high of $(2.96\pm 0.57)\times 10^{-9} \mathrm{erg\,cm^{-2}\,s^{-1}}$. At $z=0.997$, this flaring episode makes OP~313 one of the most luminous AGNs ever recorded in $\gamma$ rays. Follow-up observations with the Large-Sized Telescope prototype (LST-1) Cherenkov telescope began in December 2023 and led to the first detection of this source at VHEs, making OP~313 the most distant AGN ever detected by a Cherenkov telescope \citep{cortina2023}. Since then, the CTAO LST Collaboration has led a very intense multi-wavelength monitoring campaign of the source, which has driven the development of new analysis techniques like the one presented in this work. The detection of such a bright and distant source, coupled with the methodologies described herein, offers a unique opportunity to constrain the dynamic emission of the source and provide a good benchmarking tool to test EBL models using physically-motivated intrinsic emission models.  

In this work, we aim to validate a new analysis and data management workflow using the OGIP format\footnote{\url{https://heasarc.gsfc.nasa.gov/docs/heasarc/ofwg/docs/spectra/ogip_92_007.pdf}} \citep{1995ASPC...77..219C}, the Gamma Astro Data Formats (GADF) initiative 
\citep{2017AIPC.1792g0006D,universe7100374}, and the open-source tool \texttt{gammapy} \citep{2023A&A...678A.157D}, and provide a working prototype data archive for the source, retaining instrument response metadata for reproducibility in future analyses and interpretation of these observations. This manuscript focuses on 
demonstrating and validating a novel methodology for multi-wavelength analysis. For this demonstration we combined datasets from a small coordinated multi-wavelength follow-up campaign on OP~313 developed on March 4, 2024 (MJD60373) and March 15, 2024 (MJD60384), utilizing public data from \textit{Swift}-UVOT, \textit{Swift}-XRT, NuSTAR, and \textit{Fermi}-LAT, and data from Liverpool IO:O in the optical regime. The resulting data products, in standard OGIP/GADF formats compatible with \texttt{gammapy}, are readily available in Zenodo \citep{nievas_rosillo_2024_14033304} and  GitHub\footnote{\url{https://github.com/mireianievas/gammapy_mwl_workflow}} and described in more detail in Appendix \ref{appendix:software_description}. We defer the integration of this dataset with a larger campaign, including proprietary VHE $\gamma$-ray data from LST-1 and MAGIC, as well as additional optical, infrared, and radio results, for a separate work coordinated by the CTAO LST Collaboration. To keep the main body of the manuscript as focused as possible, we will refer always to data and results from the first observing night  on March 4, 2024 (MJD60373), and comment briefly on the results of the application of the analysis framework to the second night on March 15, 2024 (MJD60384) in Appendix \ref{appendix:second_night}.

Our proposed method involves a forward modeling approach within a unified framework for all multi-wavelength 
datasets, covering near-infrared (z-band) to high-energy (HE, $100\,\mathrm{MeV}<E<100\,\mathrm{GeV}$) $\gamma$-rays, spanning nearly 11 orders of magnitude in energy. 
The mathematical representation of this method is as follows:

\begin{equation}
\begin{split}
N (E, X) = \sum_{{\rm sources}} \mathcal{R}& (E,E_{\rm true},X) \ast \mathcal{P} (E_{\rm true},X,X_{\rm true}) \\
\ast &\left[ \mathcal{E}(E_{\rm true},X_{\rm true}) \times \phi (E_{\rm true})\right], \label{eq:general}
\end{split}
\end{equation}
where the number of expected excess event density $N(E,X)$ (in space and energy) is given as a function of the sum over a number of finite emitters of the differential spectrum $\phi (E_{\rm true})$ 
of each source times the exposure  $\mathcal{E}(E_{\rm true},X_{\rm true})$ at its location, convoluted with the PSF $\mathcal{P} (E_{\rm true},X,X_{\rm true})$, which informs about the actual distribution of the measured counts given a location in the detector/sky, possibly as a function of energy, and a redistribution or migration matrix $\mathcal{R} (E,E_{\rm true},X)$ which provides the redistribution between the true energy and the measured energy of the event, as a function of the location in the detector $X$ and the sky $X_{\rm true}$.

The manuscript is structured as follows: Section \ref{sec:motivation} details the motivation for developing a 
new method for analyzing and archiving high-level data across multiple bands. Section \ref{sec:datasets} describes 
the different instruments and procedures used to build the datasets and Instrument Response Functions (IRFs). 
Section~\ref{sec:results} presents the validation of the data formats and analysis methods, along with a 
phenomenological model of the emission of the source. The discussion of limitations, possible extensions, and future 
work is included in Section \ref{sec:discussion}, and the main results are summarized in Section \ref{sec:conclusions}. 
Wherever applicable, we use a flat $\mathrm{\Lambda}$CDM cosmology, with $\mathrm{H_0=70\,km\,s^{-1}\,Mpc^{-1}}$, 
$\mathrm{\Omega_M}=0.3$, and $\mathrm{T_{CMB,0}=2.725 \ K}$.

\section{Motivation}\label{sec:motivation}

Accurately modeling the spectral energy distribution (SED) of flaring AGNs across multiple wavelengths is a complex task due to the inherent challenges in handling diverse data types and formats. The traditional approach isolates each instrument or energy regime to `reconstruct' or estimate flux points and upper limits (ULs), followed by a reinterpretation of the same points using physically motivated models \citep[see e.g.][]{2015ApJ...815L..23A,2023A&A...670A..49M}. However, this method has several limitations.

First, storing flux points in tables or plots often results in significant information loss. This approach typically fails to account for the energy resolution of the instruments if no unfolding \citep{2017EPJWC.13711008S} is performed, a process that is inherently ill-posed. Consequently, the flux points may become correlated. Additionally, errors are commonly treated as Gaussian, which is inappropriate for instruments that operate in low-count regimes where Poisson statistics are more suitable \citep[see e.g.][]{poisson_statistics_example}. ULs are frequently represented as single values at specific confidence levels, resulting in the loss of information contained in the underlying probability distributions. This practice can lead to biased interpretations, especially when statistical fluctuations, ULs and non-detections are ignored \citep[e.g.][]{2014APh....57...30E}.

Moreover, current methods fall short in handling multiplicative models in SED flux points, such as correcting for hydrogen column density ($N_{H}$) in X-ray data, Galactic extinction in optical/UV or EBL absorption in the $\gamma$-ray regime. Typically, these corrections are applied to the reconstructed flux points, which can hinder accurate statistical correction for absorption components in the SED, particularly if the correction is applied differently for each dataset \citep[this is sometimes the case for works based on archival data samples, e.g.][]{2022MNRAS.512..137N}.

Recent advancements have explored moving away from flux points towards forward-folding techniques, which have demonstrated promising results in characterizing the EBL \citep{2019MNRAS.486.4233A} using VHE data. This forward-folding approach offers a more comprehensive statistical analysis of multi-wavelength datasets, addressing many shortcomings of traditional flux-point methods. 
One key distinction is that flux-point fitting loses access to instrument-specific information (such as energy resolution, effective area, instrumental backgrounds, and PSF) while forward-folding incorporates these details as an essential part of the analysis convolving complete emission models with the IRFs to calculate the likelihood.
By adopting the forward modeling technique, emission and absorption models can be seamlessly integrated into the analysis, enabling better handling of uncertainties in parameters such as hydrogen column density and EBL. 
Despite its advantages, forward-folding presents practical challenges, such as the need to install and configure analysis tools for each instrument and manage the orchestration of communication between them to compute the multi-instrument likelihood.

Our immediate goal is therefore to integrate high-level data from various photon-counting experiments into a unified analysis framework. To facilitate this, we propose using the standardized format for high-level data products, which ensures easier distribution of the data and instrument description, as well as reproducibility of the results. \texttt{gammapy} \citep{acero_2024_10726484}, an open-source Python package based on the popular libraries {\tt numpy} \citep{harris2020} and {\tt astropy} \citep{astropy2013,astropy2018,astropy2022,astropy_collaboration_2024_13860849}, is particularly well-suited for this task. As the official science tool for the Cherenkov Telescope Array Observatory (CTAO), \texttt{gammapy} is actively developed and compatible with both OGIP (X-ray) and GADF ($\gamma$-ray) data formats. This makes it an ideal choice for our analysis, enabling the joint analysis of different types of datasets, whether one-dimensional (1D) if only a distribution of counts as a function of energy is stored, or three-dimensional (3D) where data cubes of three dimensions (two spatial coordinates and one energy coordinate) are available. Finally, we show that even in the case of single-channel photometric datasets the proposed format is well suited to describe the data. 

Furthermore, \texttt{gammapy} allows the incorporation of physically motivated models through external emission libraries --- including synchrotron, inverse Compton, absorption components such as EBL absorption --- or even extend it with models from the X-ray library {\tt sherpa} \citep{sherpa1,sherpa2,doug_burke_2024_13909532} to add hydrogen absorption, and interstellar extinction. Public radiation processes modeling codes like \texttt{agnpy} \citep{agnpy,nigro_2023_7633553}, \texttt{jetset} \citep{jetset1,jetset2,jetset3}, and \texttt{naima} \citep{naima} offer a more accurate representation of underlying physical processes compared to traditional functional or empirical models available in standard X-ray analysis packages like \texttt{xspec} \citep{xspec} and \texttt{sherpa}. This capability allows for a more accurate and statistically robust comparison of different models.

The proposed workflow can be compared with the Multi-Mission Maximum Likelihood framework \citep[3ML,][]{2015arXiv150708343V,2021zndo...5646954B}, a versatile tool designed to facilitate multi-wavelength analysis by integrating various software packages tailored for different types of observational data \citep[see e.g.][]{2023ApJ...942...96A,2024MNRAS.529L..47K}.  However, {\tt 3ML} requires a complex setup involving the installation and configuration of multiple software tools within a common environment to `export' the likelihood functionality for each instrument. This integration involves ensuring seamless communication between tools such as {\tt xspec} for X-ray data, {\tt Fermipy} or \texttt{gammapy} for $\gamma$-ray analysis, and other domain-specific software for optical photometric data. In contrast, while our method requires the preliminary construction of IRFs, it provides a unified data format and analysis workflow that supports the integration of multi-wavelength data without the need for multiple software environments. By design, the proposed methodology simplifies the analysis setup, enhances portability, and improves reproducibility compared to the arguably more cumbersome process required by {\tt 3ML}.

In the following sections, we will detail the components of our methodology, and describe the dataset and IRF generation for the case of study of the flaring blazar OP~313. Our proposed framework aims to enhance the accuracy of SED modeling and streamline the analysis process across diverse astrophysical datasets and improve the reproducibility of the results. This work focuses on the technical implementation and presents validation tests in comparison to the standard analysis tools and methods used for each instrument.

\section{Dataset construction}\label{sec:datasets}

\subsection{Fermi-LAT data reduction}

The Large Area Telescope \citep[LAT,][]{atwood2009} onboard the \textit{Fermi} satellite is a space-based, pair-conversion telescope surveying the sky in the high-energy $\gamma$-ray band ($\lesssim 100\,$MeV to $\gtrsim 100\,$GeV). It covers $20\%$ of the sky instantaneously and surveys the entire sky approximately every $\sim 3$ hours, making it an excellent instrument for studying transient phenomena, including AGN flares.

\subsubsection{Data reduction}

We collected \textit{Fermi}-LAT \texttt{Pass 8 SOURCE} class events with energies larger than $100\,$MeV within a region of interest (ROI) of $15\,$deg radius around the OP~313 position from October 1, 2023, to April 1, 2024 (183 days). Nightly binned analyses, centered at 00:00 UTC, were performed with the \textit{Fermi} Science Tools \citep{fermitools}, using \texttt{enrico} \citep{sanchez2015} as a high-level wrapper to manage the individual jobs. We utilized the latest available version of LAT's IRFs ({\tt P8R3\_SOURCE\_V3}) and applied a conservative zenith cut of $90\,$deg to avoid Earth's limb contamination and a {\tt DATA\_QUAL==1 \&\& LAT\_CONFIG==1} filter to ensure that only good quality events were analyzed, following the recommendations in Cicerone\footnote{\url{https://fermi.gsfc.nasa.gov/ssc/data/analysis/documentation/Cicerone/}}.

The sky model for the analysis was generated using the \textit{Fermi}-LAT 14-Year Point Source Catalog \citep[4FGL-DR4][]{abdollahi2022, ballet2023}, incorporating sources within a $20\,$deg radius around OP~313. Because we only integrate $24\,$h per analysis, we adopted a simple power-law (PWL) spectral model for OP~313, including the EBL absorption model from \cite{dominguez2011}. Positions and extensions of all sources from the 4FGL-DR4 were kept fixed, leaving only the spectral parameters of OP~313 (normalization and spectral index) and the normalization of the isotropic background component as free parameters. Spectral parameters for weak sources ($<10\,\sigma$ significance) and distant sources ($>5\,$deg) were kept fixed, except for normalizations. 

\subsubsection{Long-term analysis of the high-energy flare}

\begin{figure*}
\begin{subfigure}[b]{.65\linewidth}
    \centering
    \includegraphics[width=1\linewidth]{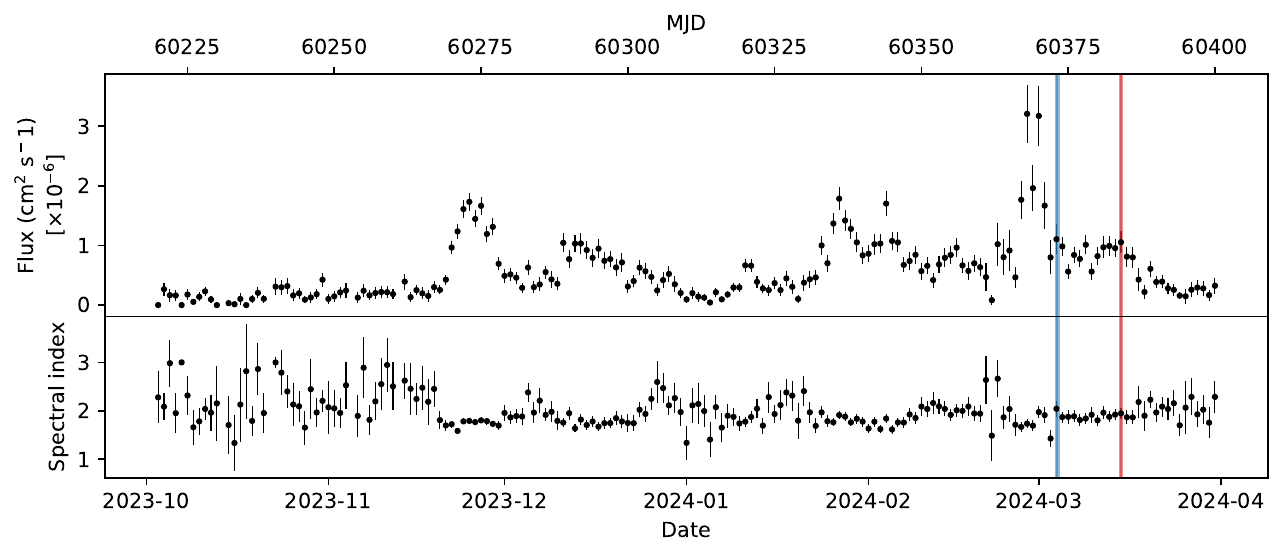}
\end{subfigure}
\begin{subfigure}[b]{.285\linewidth}
\centering
\includegraphics[width=1\linewidth]{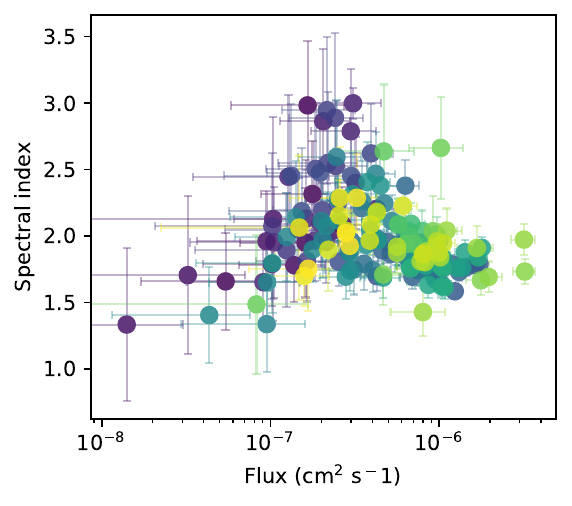}
\end{subfigure}
    \caption{\textbf{Left:} \textit{Fermi}-LAT light curve (1-day binning) of OP~313 above 100~MeV during the major 2023-2024 flare episode, for integral flux and spectral index. Blue and red bars mark the two nights with NuSTAR observations that are the focus of this study. \textbf{Right:} Spectral index as a function of the integral flux for each night over six months of LAT data. The density of points is color-coded, with yellow points showing larger number density of nights and purple-blue tones indicating more isolated bins.
    }
    \label{fig:Fermi_LC}
\end{figure*}

In addition to focusing on two specific nights with comprehensive multi-wavelength coverage, we examined the broader activity of OP~313 to contextualize its emission during these nights within a longer-term flare that persisted over several months in 2023-2024.  Figure~\ref{fig:Fermi_LC} shows the evolution of the nightly flux above $100\,$MeV for OP~313, with 24-hour bins centered at midnight UTC. It also depicts the evolution of the spectral index in the \textit{Fermi}-LAT energy band and its correlation with the estimated flux, hinting that during high emission states, the spectrum of OP~313 becomes harder, with an index in the LAT band smaller than $\sim 2$. Figure~\ref{fig:Fermi_LC} shows that the source entered an intermediate $\gamma$-ray activity plateau phase following a very strong flare at the end of February. 

\subsubsection{Dataset and IRFs}

\begin{figure*}
    \centering
    \includegraphics[width=0.9\linewidth]{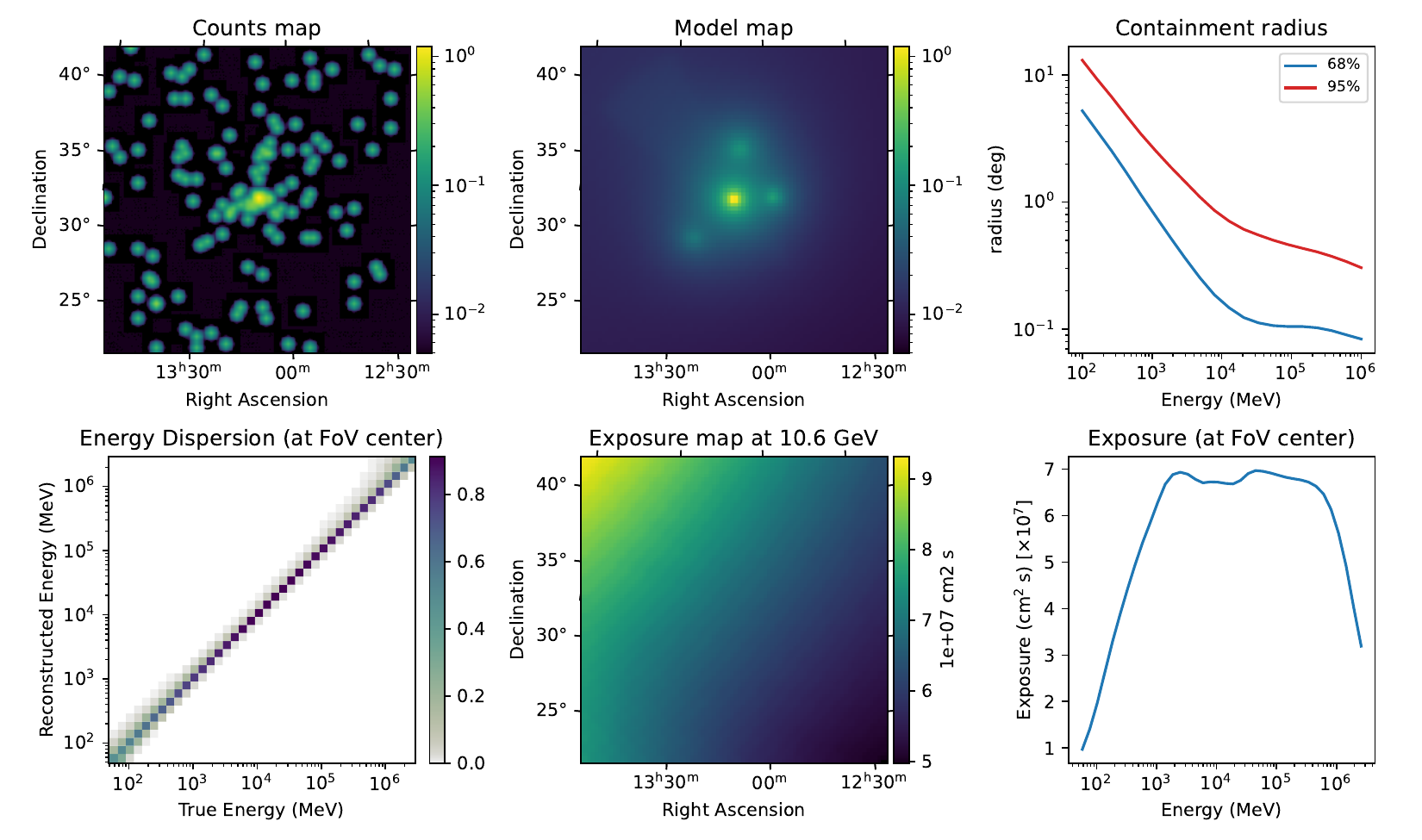}
    \caption{{\it Fermi}-LAT dataset description for the first selected night. From top left to bottom right: \textbf{(a)} Integral counts map, smoothed with a Gaussian kernel of $0.2^\circ$. \textbf{(b)} Initial model map, convolved with the IRFs ({\tt npred} map). \textbf{(c)} $68\%$ and $95\%$ containment radius as a function of energy. \textbf{(d)} Energy dispersion (reconstructed energy as a function of true energy). \textbf{(e)} Exposure map at $10.6 \,$GeV. \textbf{(f)} Exposure at the FoV center as a function of true energy.}
    \label{fig:Fermi_IRFs}
\end{figure*}

For {\it Fermi}-LAT, the dataset is managed as a 3D \texttt{MapDataset} in \texttt{gammapy}, which includes spatial (two coordinates) and energy (one coordinate) dimensions. Figures \ref{fig:Fermi_IRFs} shows 2D and 1D projected representations of the different parts giving shape to the Dataset. From first row, left, to second row, right, the following information is shown:

\begin{enumerate}[a]
    \item {Integral Counts Map}: a 2D representation of the underlying `counts cube' summed over the entire energy range and smoothed with a Gaussian kernel of $0.2$ deg to enhance visibility of sources.
    
    \item {Initial Model Map} convolved with the IRFs to produce the predicted counts ({\tt npred} map) and again integrated over the entire energy range just for visualization purposes. The original 3D version of this Model is fit to the data (`counts cube') by \texttt{gammapy}.

    \item {Containment Radius:} 1D representation of the PSF corresponding to $68\%$ and $95\%$ of the events contained as a function of energy, indicating the spatial resolution of the LAT. As opposed to the native analysis, {\tt Fermitools}, our conversion to \texttt{gammapy} format assumes a non-varying PSF. For the analysis of a bright point-like source in a $20^\circ$ region, the impact of spatial variations of the PSF of LAT is negligible.

    \item {Energy Dispersion:} Distribution of reconstructed energies versus true energies, providing information on how accurately the instrument measures the energy of $\gamma$-ray events and energy bias (significant at the lowest energies, $E\lesssim 100\,$MeV).

    \item {Exposure Map:} 2D projection of the exposure (effective area multiplied by livetime) at $30.6\,$GeV, illustrating the spatial variation in exposure across the FoV.

    \item {Exposure at the FoV center:} 1D projection of the exposure as a function of true energy at the center of the field of view, showing the energy dependence of LAT's sensitivity.
\end{enumerate}

\subsection{NuSTAR data reduction}

The Nuclear Spectroscopic Telescope Array \citep[NuSTAR,][]{harrison2013} is a focusing high-energy X-ray telescope operating in the 3 to 79 keV energy range. NuSTAR is a telescope array consisting of two twin modules (NuSTAR~A and NuSTAR~B), each made of four Cadmium-Zinc-Tellurium detectors.

The standard data products provided to the observer are similar to other X-ray and gamma-ray space telescopes and consist of an event list with the reconstructed position, energy and time of arrival of arrival of each detected photon.

The typical data reduction procedure used in X-ray astronomy is to define a spatial region for spectral extraction and an OFF region to estimate the spectral background. 
The data reduction and first stages of the analysis were performed with {\tt nustardas v2.1.2}. The data were obtained from the quick-look area\footnote{\url{https://nustarsoc.caltech.edu/NuSTAR_Public/NuSTAROperationSite/Quicklook/Quicklook.php}} for observation IDs 91002609002 (night of March 4th, 2024) and 91002609004 (night of March 15th, 2024). We ran {\tt nupipeline} to process the data through Stage 1 (data calibration) and Stage 2 (data screening) for each of the nights and each of the telescope modules. 

\begin{figure}
    \centering
    \includegraphics[width=1\linewidth]{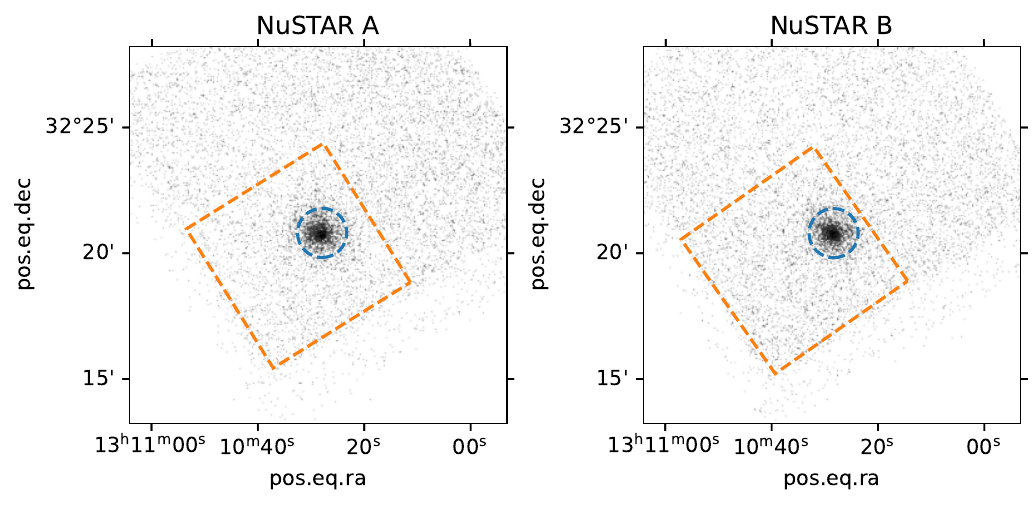}

    \caption{NuSTAR A/B counts map for the first observation night, together with the signal extraction region (blue dashed circle) and background extraction region (orange dashed box minus the signal extraction circle).}
    \label{fig:NuSTAR_countsmap}
\end{figure}

In a second step, we ran {\tt nuproducts} (Stage 3) over the Level 2 event files to extract the integrated bandpass images ($3-79\,$keV), the source spectrum and light curves. Each NuSTAR module and sector has slightly different properties, therefore the background extraction regions were optimized for each module in shape and location to maximize background statistics, but keeping it within the limits of the same sector in which the source is located (see Fig. \ref{fig:NuSTAR_countsmap}). From this point, we focus only on the spectral analysis of the Level 3 data produced by {\tt nuproducts}, consisting of: \textbf{(i)} {\tt PHA} files (energy spectra) for the source and background; \textbf{(ii)} {\tt ARF} file (Ancillary Response File), which contains the effective area; and \textbf{(iii)} {\tt RMF} file (Response or Redistribution Matrix File), containing the energy migration matrix, a 2D-array that gives the probabilities of assigning the photon to a given PHA channel for each input photon energy.

\subsubsection{Background model}

Two analyses of the spectral data were considered, each with different background assumptions. The first analysis uses the ON-OFF method with W-statistics (W-statistics, wstat\footnote{Note that in xspec, no name distinctions are made between cstat and wstat, see \url{https://heasarc.gsfc.nasa.gov/docs/xanadu/xspec/manual/XSappendixStatistics.html}},  or Poisson data with a Poisson background). In this approach, the background is estimated measuring the number of events in the OFF region (see Fig.\ref{fig:NuSTAR_countsmap}), assuming that it is the same as in the source region once correcting for the region size differences.
This method is straightforward but cannot account for any spatial variation of the background (e.g. instrumental lines).
In addition, it is statistically limited, as the W-statistic estimator can produce inaccurate results in low-count regimes, particularly if the background reaches zero counts in a given energy bin. To mitigate these issues, we re-binned the data into coarser energy bins (by a factor of 32) to increase the statistics per bin.

The second approach involves using a background model for NuSTAR data. We employed the Cosmic X-ray Background (CXB) simulation tool \texttt{nuskybgd} \citep{2014ApJ...792...48W} to estimate the background (using C-statistics, cstat, or Poisson data with a background model). Compared to a directly measured background via the OFF background method, the predicted background from \texttt{nuskybgd} effectively addresses issues related to very low statistics at high energies. At the cost of the complexity added by \texttt{nuskybgd}, this approach retains better spectral resolution while fully utilizing the energy range of NuSTAR, and accounts for the spatial variation of instrumental background.
A detailed comparison of the impact of the choice of the background method is presented in Appendix \ref{appendix:background_nustar}.

\subsubsection{Dataset and IRFs}

\begin{figure}
    \centering
    \includegraphics[width=1\linewidth]{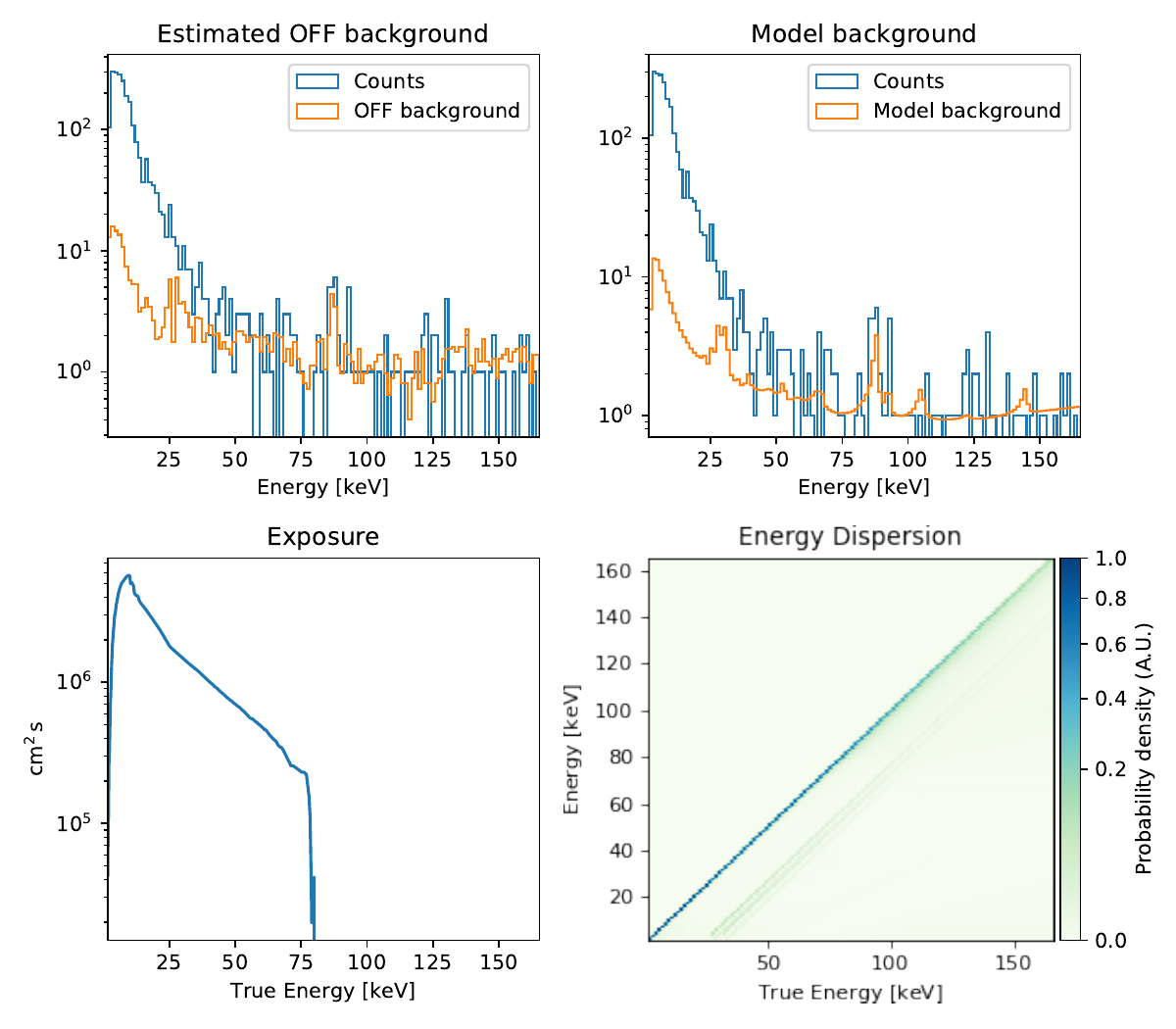}
    \caption{\texttt{gammapy} dataset representation of NuSTAR A for night 1 is shown here. NuSTAR B for this observation and NuSTAR A/B for night 2 are similar. From top left to bottom right: \textbf{(a)} Estimated distribution of background events from the OFF rectangular region of Figure \ref{fig:NuSTAR_countsmap} and the ON event distribution from the signal extraction region. \textbf{(b)} Estimated distribution of background events from {\tt nuskybgd}. ON event distribution identical to the previous case. \textbf{(c)} NuSTAR A exposure (effective area $\times$ livetime) as a function of true energy. \textbf{(d)} NuSTAR A energy dispersion matrix. 
    }
    \label{fig:NuSTAR_peek_rebin}
\end{figure}

An overall view of the datasets for the NuSTAR~A module on the first of the two available nights, with both the Poisson measured background (\texttt{gammapy}'s  {\tt SpectrumDatasetOnOff} object) and the one with {\tt nuskybgd} background model (\texttt{gammapy}'s {\tt SpectrumDataset} object), is shown in Fig. \ref{fig:NuSTAR_peek_rebin}. 

\subsection{Swift-XRT data reduction}

The X-Ray Telescope \citep[XRT,][]{burrows2004} onboard the \textit{Swift} satellite is an X-ray CCD imaging spectrometer designed to measure the position, spectrum, and brightness of astronomical objects in the range of $0.2,$ to $10,$ keV, with a dynamic range spanning more than seven orders of magnitude in flux. XRT has an effective area of approximately $135\,\mathrm{cm^2}$ at 1.5 keV and a detection sensitivity of $2\times10^{-14}\, \mathrm{erg\,cm^{-2}\,s^{-1}}$ in $10\,$ ks of observing time.

The FoV is $23.6 \times 23.6$ arcmin, with an angular resolution of $18$ arcsec (Half-Energy Width, HEW). XRT operates in two readout modes: Windowed Timing (WT) and Photon Counting (PC). The WT mode bins 10 rows into a single row, focusing on the central 200 columns to improve time resolution to $1.7\,$ ms (ideal for bright sources). The PC mode retains full imaging and spectroscopic capabilities but with a slower time resolution of $2.5$\,s, and is limited by pileup effects at high count rates \citep[$\gtrsim 0.6\,$ c/s,][]{2006ApJ...638..920V}. OP~313 is a relatively weak X-ray source with a modest count rate of $\lesssim 0.2\,$ c/s, well within the capabilities of XRT in PC mode; therefore, this work focuses exclusively on this mode.

\subsubsection{Data reduction}

{\it Swift}-XRT data for OP~313 were downloaded from the UK Swift Science Data Centre\footnote{\url{https://www.swift.ac.uk/swift_portal/}}. Of all available data from the telescope, this work focuses on data collected during the nights of March 4th, 2024 and March 15th, 2024, with Swift observation IDs 00036384074 ($2.96\,$ks) and 00089816002 ($3.16\,$ks), respectively. We used \texttt{Heasoft} 6.32.1 and \texttt{xrtdas v3.7.0} for data reduction. 
Spectra were extracted using 
\texttt{xrtproducts} for a fixed source extraction region of $40"$  radius, and a surrounding annular region of $80"$ and $300"$ radii for the background (see Fig.~\ref{fig:SwiftXRT_peek_rebin}). This process results in two separate PHA files, along with the corresponding ARF. The RMF file was downloaded from HEASARC's calibration database (CALDB). 

\subsubsection{Dataset and IRFs}

\begin{figure}
    \centering
    \includegraphics[width=1\linewidth]{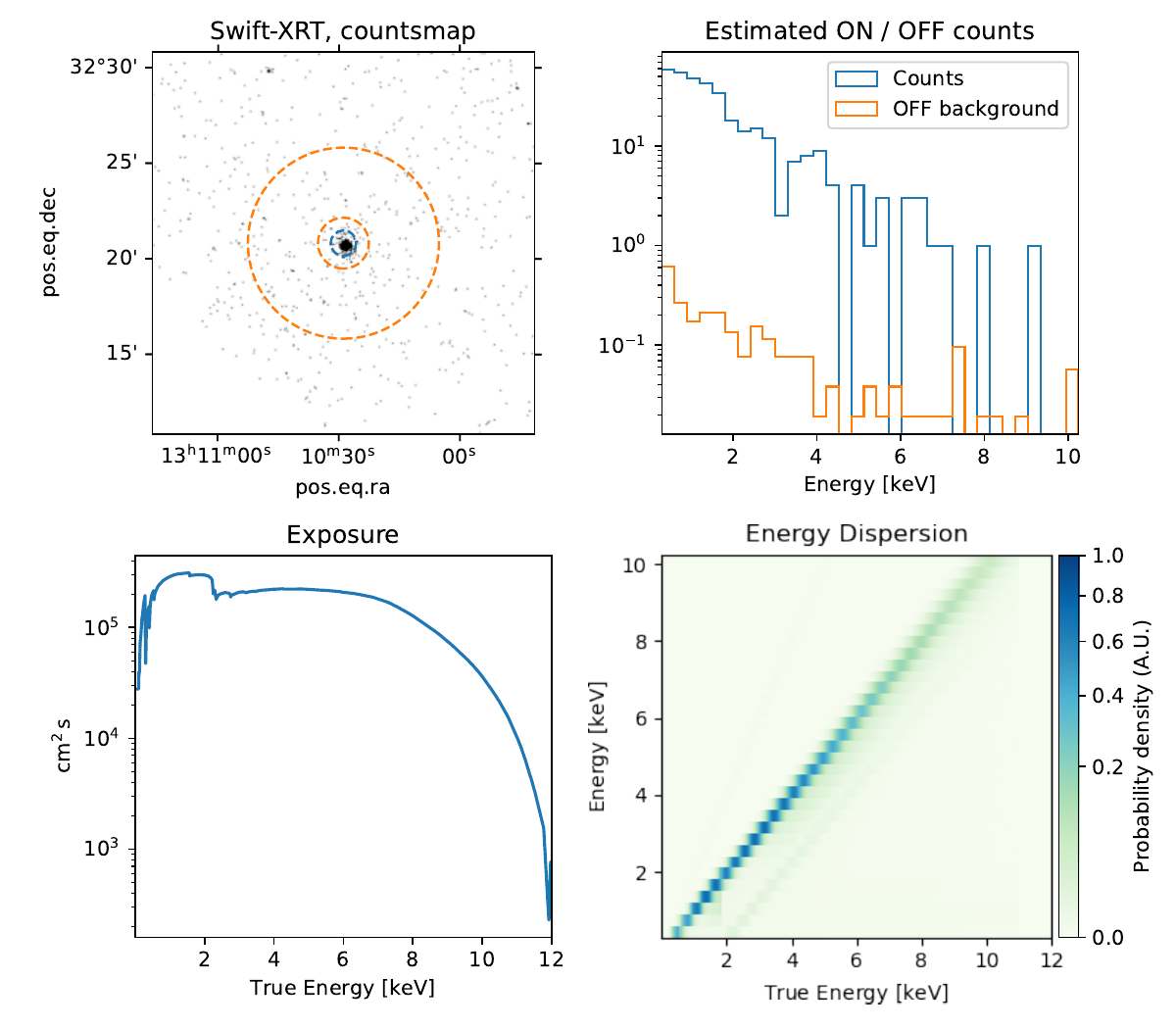}
    \caption{\textit{Swift}-XRT dataset representation after energy bin grouping to avoid background spectra dominated by zeros. \textbf{(a)} \textit{Swift}-XRT image of OP 313. Blue circle and orange annulus show the signal and background extraction regions, respectively. \textbf{(b)} Distribution of events for both regions as a function of energy, corrected for the different region sizes. \textbf{(c)} \textit{Swift}-XRT exposure (effective area $\times$ livetime) as a function of true energy. \textbf{(d)} \textit{Swift}-XRT energy dispersion matrix (also known as redistribution matrix or migration matrix).}
    \label{fig:SwiftXRT_peek_rebin}
\end{figure}

The source and background PHA files, together with the RMF and ARF, were used to generate \texttt{gammapy}'s \texttt{SpectrumDatasetOnOff} objects, represented in Fig. \ref{fig:SwiftXRT_peek_rebin}. Following the discussion for NuSTAR on background statistics, we re-binned the spectral data into coarser energy bins with 30 channels each, to avoid zero counts in the background bins and therefore preserving the accuracy of the W-statistics assumption.

\subsection{Swift-UVOT data reduction}\label{sec:datasets:uvot}

The Ultraviolet/Optical Telescope \citep[UVOT,][]{roming2005} onboard the \textit{Swift} satellite is a diffraction-limited $30\,$cm modified Ritchey-Chr\'etien telescope with a limiting magnitude of $22.3$ in $1\,$ ks exposures. The UVOT detector, identical to XMM-Newton's microchannel-plate intensified CCD, operates in photon counting mode over the range of $170$ to $650\,$ nm with 7 filters (V, B, U in the visible band; UVW1, UVM2, UVW2 in the UV band; and a white or full-passband filter).

\subsubsection{Data reduction}

Data reduction for \textit{Swift}-UVOT is performed using \texttt{HEASoft} v6.32.1. 
After downloading the target images from HEASARC’s repository, source extraction regions were defined as circular apertures. We used an exceptionally large radius of $25"$ as one {\it Swift} gyroscope gradually degraded since July 2023 until failing in March 2024, increasing the likelihood of star tracker ``loss of lock'' after long slews \citep{gcn1,gcn2}. The effect was particularly strong in the second night. The source regions are initially centered on the nominal position of OP~313 and fine-tuned based on centroid calculations. A background extraction region is then defined as an annulus with an inner radius of $40"$ and an outer radius of $100"$, centered on the source region. 
An example of the source and background regions, overlaid on top of the cutout of the \textit{Swift}-UVOT images for different filters, is shown in Figure~\ref{fig:UVOT_images}. OP~313 is a distant, strongly Doppler-boosted source that appears point-like in most energy bands, including optical and UV. Consequently, UVOT data are analyzed using the same strategy used for XRT and NuSTAR. The result is the production of PHA source and background files (containing the integrated number of counts) for each filter. UVOT Response files (RSP) are downloaded directly from CALDB.

\subsubsection{Dataset and IRFs}

\begin{figure}
    \centering
    \includegraphics[width=0.9\linewidth]{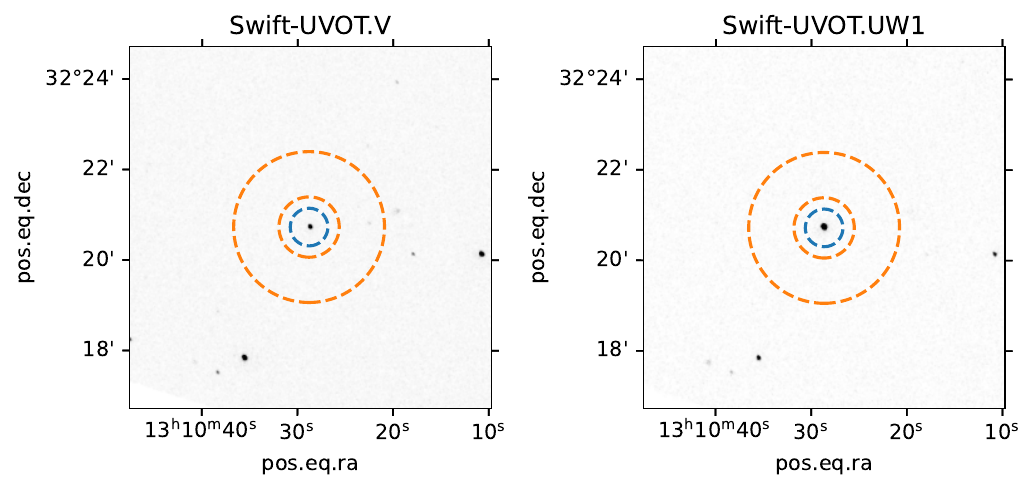}
    \includegraphics[width=0.9\linewidth]{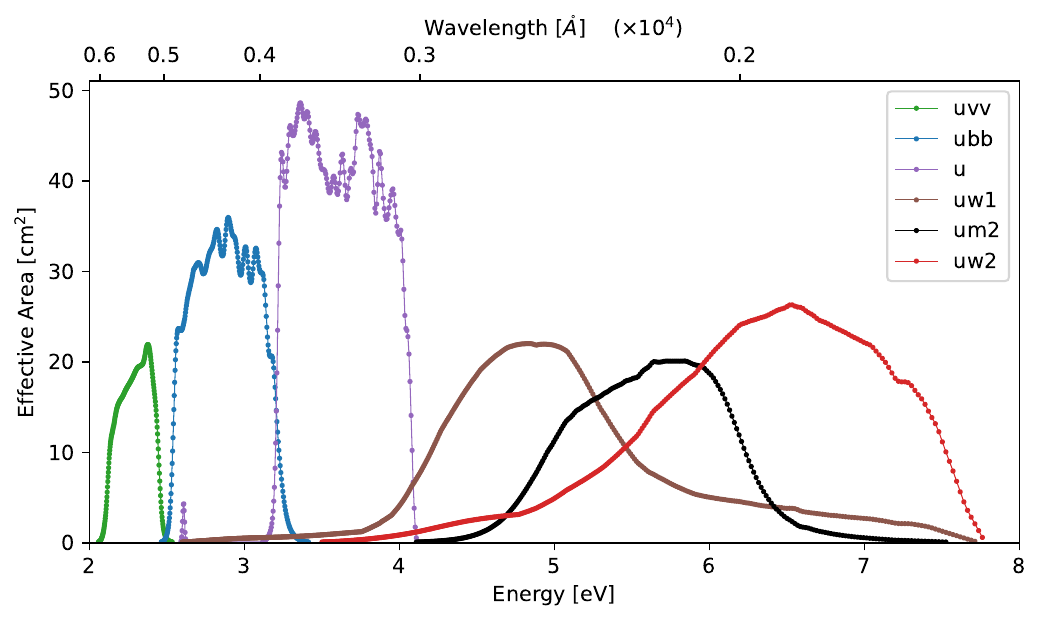}
    \caption{\textbf{First row:} \textit{Swift}-UVOT images in two of the photometric filters available for the first selected night, indicated by dashed blue circles (signal) and orange annuli (background), respectively. \textbf{Second row:} Effective collection area for each of the available \textit{Swift}-UVOT filters.}
    \label{fig:UVOT_images}
\end{figure}

In imaging mode (i.e., without using the grisms) and for point-like sources, we treat UVOT as a single-channel detector with a response determined by the selected photometric filter. The RSP files provide effective area information for each filter (see Figure \ref{fig:UVOT_images}, second row). A dummy redistribution matrix is generated on the fly to match \texttt{gammapy}'s expected format definition. These redistribution matrices consist of a row of constant values, indicating that all photon energies contribute to the single-channel detector.

\subsection{Liverpool-IO:O data reduction}

The Liverpool Telescope \citep[LT,][]{steele2004}, located on the Canary island of La Palma, is a fully robotic 2-meter telescope operated by Liverpool John Moores University. Its robotic capabilities allow for seamless transitions between different observing programs and instruments, which is crucial for time-sensitive observations of transient phenomena such as $\gamma$-ray bursts and supernovae. The IO:O
\citep[Infrared-Optical:Optical, see][]{barnsley2016} camera on the LT is equipped with a charge-coupled device (CCD) and a range of photometric filters. This camera includes a complete Sloan filter set: $u$, $g$, $i$, $r$, $z$, making it a valuable lower-energy complement to the \textit{Swift}-UVOT.

We developed a custom data analysis pipeline using the {\tt photutils}\citep{larry_bradley_2022_6825092} Python package to process LT IO:O calibrated images through differential aperture photometry. This pipeline reduces each image into a {\tt SpectrumDatasetOnOff} compatible format in \texttt{gammapy}, calculating signal and background counts in ON and OFF regions, respectively. The exposure or instrument response is estimated with the same images, using a differential photometry technique consisting on measuring the flux of several calibration stars in the FoV and comparing it to their known flux densities in reference catalogs.

\subsubsection{Calibration stars}

For each image, the pipeline utilizes the WCS information in the FITS header to identify potential calibration stars. 
These stars are matched with a clean star sample from the SDSS-DR18 catalog \citep{almeida2023}, downloaded using \texttt{astroquery}\citep{2019AJ....157...98G} with a custom SQL query to select all bright stars (m$_{u,g,r,i,z}<\,$21) in the FoV of the LT images, except: those with non-clean flags, stars close to the edges of the SDSS images, stars affected by cosmic rays, blended stars with non-robust photometry, those with large PSF magnitude errors or saturated pixels. An additional query to the Pan-STARRS PS1 catalog \citep{chambers2016, flewelling2020} is performed to remove variable stars, as this catalog provides a measurement of the star magnitude dispersion (in bands $g$, $r$, $i$ and $z$) for different observations of the same field. 
For each of the surviving stars (9 for our images), we assemble an \texttt{astropy} Table object containing the coordinates, magnitudes in the five selected bands and errors in the magnitudes. The magnitudes reported in the SDSS-DR18 are in the AB system, therefore conversion to flux densities, $F_\nu$, is given by: 

\begin{equation}
m_\mathrm{AB} = -2.5 \log_{10} \left[ \frac{F_\nu}{\mathrm{erg\, cm^{-2}\, s^{-1}\, Hz^{-1}}} \right] -48.60
\end{equation}

\subsubsection{Exposure estimation}

The expected number of excess counts \( N_{\text{src,tot}} \) for an ON/OFF dataset in a given energy window can be expressed, following Eq. \ref{eq:general}, in terms of the exposure \( \mathcal{E}(E_{\rm true}) \) and the differential spectrum \( \phi (E_{\rm true}) \) as:

\begin{equation}
N_{\text{src,tot}} = \int_{E} \mathcal{R} (E,E_{\rm true}) \ast \left[\mathcal{E}(E_{\rm true}) \times \phi (E_{\rm true})\right] \, dE_{\rm true}. \label{eq:Ntot}
\end{equation}

For optical ground-based instruments, characterizing \( \mathcal{E}(E_{\rm true}) \) is complex due to the need to account for both optical elements and atmospheric effects, the latter particularly hard to assess and only accounted for as an energy-averaged scaling factor of the exposure which is derived directly from star photometry. We approximate the energy-dependence of the exposure (effective area multiplied by livetime, as per \texttt{gammapy} definition) by the detector throughput curves provided in \cite{barnsley2016}, consisting of the product of filter transmission times cryostat transmission times CCD quantum efficiency (see Figure \ref{fig:Liverpool_images}). This approximation assumes that the photometric band is narrow enough for the source spectrum to be approximated by the photon density \( F_\nu / \nu_{c} \), where \( \nu_{c} \) is the effective frequency of the photometric band and \( F_\nu \) is the flux density. Rewriting Eq.~\ref{eq:Ntot} in terms of wavelength, we get:

\begin{equation}
N_{\text{src,tot}} \simeq \frac{F_\nu \bar{\mathcal{E}}}{h \lambda_c} \int_{\lambda} T(\lambda) \, d\lambda, \quad \text{where}\  \lambda_c = \frac{\int_{\lambda} T(\lambda) \lambda \, d\lambda}{\int_{\lambda} T(\lambda) \, d\lambda}. \label{eq:Ntot_simplified}
\end{equation}

For each image, multiple stars are used to combine exposure estimates into a single value using a weighted average. The weights are the inverse of the estimated exposure variance. The final error on the exposure is obtained by adding the statistical uncertainty from the weighted average in quadrature to the standard deviation of all measurements:

\begin{equation}
    \langle\bar{\mathcal{E}}\rangle = \frac{\sum_k \frac{\bar{\mathcal{E}}_k}{\sigma_{\bar{\mathcal{E}}_k}^2}}{\sum_k \frac{1}{\sigma_{\bar{\mathcal{E}}_k}^2}}, \\
    \Delta \langle\bar{\mathcal{E}}\rangle = \sqrt{ \frac{1}{\sum_k \frac{1}{\sigma_{\bar{\mathcal{E}}_k}^2}} + \sum_k (\bar{\mathcal{E}}_k - \langle\bar{\mathcal{E}}\rangle)^2 }.
\end{equation}

The exposure is then defined as \( \mathcal{E}(E) = \langle\bar{\mathcal{E}}\rangle \times T (E) \), where \( T(E) \) is interpolated from the known detector throughput dependency on wavelength, $T(\lambda)$.

\subsubsection{Dataset and IRFs}

\begin{figure}
    \centering
    \includegraphics[width=0.95\linewidth]{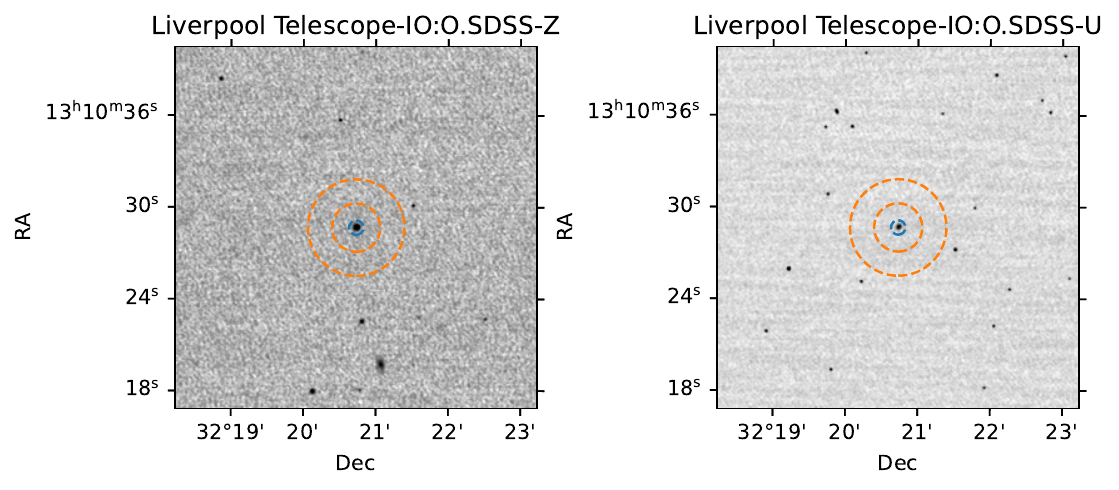}
    \includegraphics[width=0.95\linewidth]{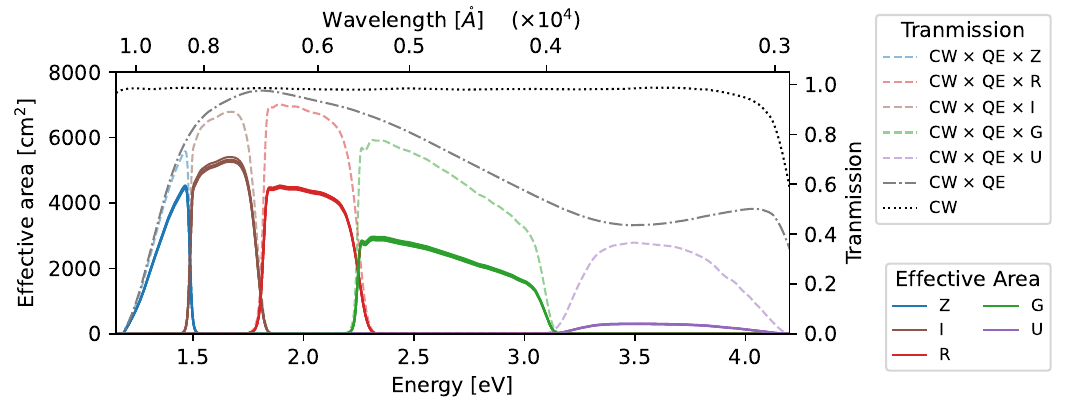}
    \caption{\textbf{Top row:} LT IO:O images in two of the photometric filters available for the first selected night, with the signal and background extraction regions represented by blue and orange dashed circles and annuli, respectively. \textbf{Bottom row:} Throughput for each of the available LT IO:O bands (transmission of the filter multiplied by the cryostat window transmission, CW, and the quantum efficiency of the CCD, QE) as dashed curves, and for CW and CW$\times$QE alone. For completeness, the reconstructed effective area for each calibration star and each filter are shown as solid colored curves.}
    \label{fig:Liverpool_images}
\end{figure}

For the LT IO:O observations of OP~313, we constructed \texttt{SpectrumDatasetOnOff}-style datasets, integrating photons over a region in the sky where the source is located for each filter (see Figure~\ref{fig:Liverpool_images}). This approach resulted in 1D datasets for each image, losing spatial information. The exposure is estimated using the method described earlier, with the RMF generated following the same process as in UVOT, consisting of a row of constant values due to the single-channel nature of each image.

In contrast to space-borne photon-counting instruments, the flux calibration relies on differential photometry using calibration stars within the FoV. Variations in star spectra introduce systematic uncertainties in the effective areas. In a classical analysis method, based on flux point generation, these systematics are commonly added in quadrature to the statistical uncertainty. 

In the forward-folding approach described here, an alternative method is utilized to include normalization models for each dataset or each filter (e.g., using a \texttt{PiecewiseNormSpectralModel} with normalization factors defined at filter-center energies). Gaussian priors are applied to account for the measured systematic uncertainties, with the widths of the Gaussian priors set to the standard deviation of the relative errors in the exposure with respect to the calculated mean. These relative errors range from less than $0.1\,$mag for the $z$- and $i$-bands to more than $0.3\,$mag for the $u$-band, where fewer reference stars are detectable. A more detailed discussion on the practical differences between the two methods and the effect of adding the systematic in quadrature to the statistical uncertainty is provided in Appendix \ref{appendix:optical_photometry_systematics}.

\section{Results}\label{sec:results}

\subsection{Summary of observational data and analysis}\label{sec:data_summary}

We present the results from the broadband emission analysis of OP~313, spanning from optical to high-energy $\gamma$ rays. The instruments used, listed in decreasing order of energy coverage, are: {\it Fermi}-LAT, NuSTAR, {\it Swift}-XRT, {\it Swift}-UVOT, and the LT with the IO:O camera. 
Our analysis aims to interpret the data in a multi-wavelength context and evaluate the capabilities of our new multi-instrument data management workflow. Before testing broadband modeling of the joint dataset, we carried out a technical validation of the analysis of data for each instrument separately, comparing in each case the spectral reconstruction of \texttt{gammapy} against that of the analysis tools native to each instrument. The details of this validation can be found in Appendix \ref{appendix:validation}.
This section is divided into three parts:

\begin{enumerate}
    \item {Phenomenological Broadband Emission Model:} We define the model for the broadband emission of OP~313, which also serves as a benchmark for the analyses.
    \item {Comparison of Fitting Approaches:} We compare the results obtained from the canonical flux point fitting method with those from the forward-folding approach.
    \item {Physical Interpretation of Emission:} We discuss the physical implications of the emission model and its consistency with the observed data.
\end{enumerate}

\begin{figure}
\centering
    \includegraphics[width=\linewidth]{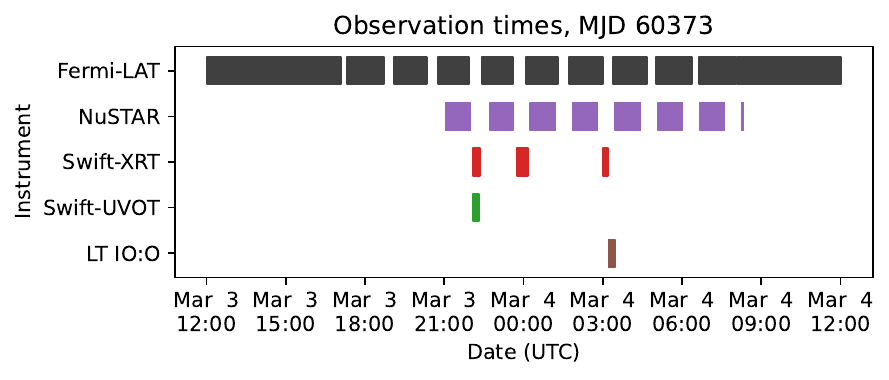}
    \caption{Temporal coverage of the first observation night.} 
    \label{fig:dataset_gti}
\end{figure}

\subsection{Broadband emission modeling}\label{sec:model}

\subsubsection{Emission components}

To model the broadband emission from OP~313, we employed a phenomenological approach using two power-laws with exponential cut-offs with a fixed index of $\alpha=1$:

\begin{equation}
    \phi_{i}(E) = \phi_{0,i} \left(\frac{E}{E_{\text{ref},i}}\right)^{-\Gamma_{1,i}} \left(-(\lambda E)^\alpha \right), \quad i\in\{LE,HE\}.
\end{equation}

In this model, $\phi_{LE}(E)$ represents the synchrotron emission (low energy component), and $\phi_{HE}(E)$ represents the inverse Compton emission (high energy component) in a simplified leptonic scenario. The $\lambda$ parameter refers to the inverse cut-off energy.

\subsubsection{Absorption components}

Absorption models were applied as follows:

\begin{itemize}
    \item {ISM extinction}: Relevant for optical and especially UV energies. 
    Given OP~313's Galactic latitude of $+83.34^\circ$ and a color excess of $E_{B-V} \approx 0.0125$ \citep{1998ApJ...500..525S}, interstellar medium (ISM) extinction is minimal below $10$ eV. At $\sim 5\,\text{eV}$ (UVOT's UVM2 filter), the transparency is approximately 90\% \citep{1989ApJ...345..245C}. 
    \item {Hydrogen I absorption}: Relevant for X-ray energies below $\sim 1\,\text{keV}$. Parameterized using the neutral hydrogen (HI) column density \cite{2000ApJ...542..914W}, which is $1.25\times 10^{20}\,\text{cm}^{-2}$ for OP~313 \citep{2016A&A...594A.116H}.
    \item {EBL absorption}: Significant at energies above $\sim 10\,\text{GeV}$. We use the model from \cite{saldana-lopez2021}, which can be regarded as an update over \cite{dominguez2011}.
\end{itemize}

\subsection{Forward folding fitting method}\label{sec:forward_folding}

We employed two distinct fitting methodologies to analyze the data from the participating instruments:

\begin{enumerate}
    \item {Flux Point Fitting}: This traditional method fits the model to observed flux points derived from analyzing each instrument's data separately. Although it simplifies the fitting and can be computationally faster, it is observed to be more sensitive to the choice of initial parameters. Additionally, this approach may introduce errors due to differences in flux estimation and instrument calibration methodologies. It also overlooks upper limits and correlations between flux points. 

    \item {Forward Folding Technique}: This method fits the model directly to the counts or event data from the instruments, convolving with the IRFs in each iteration. It allows incorporating background components and full instrumental effects. 
\end{enumerate}

\begin{figure}
    \centering
    \includegraphics[width=1\linewidth]{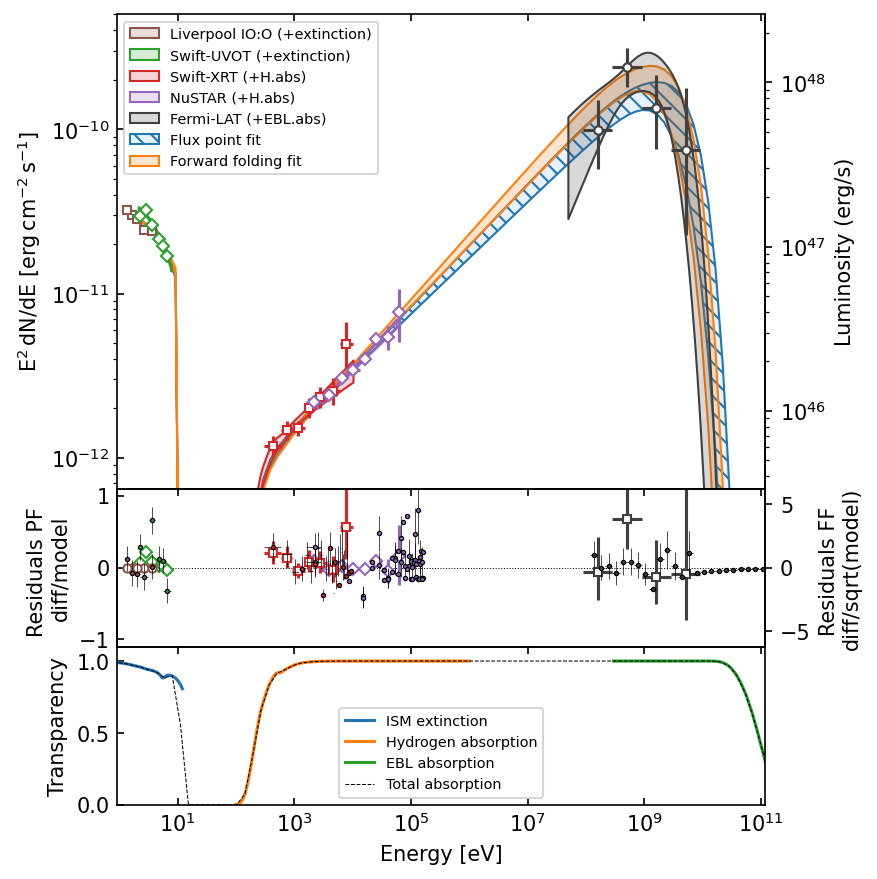}
    \caption{Joint multi-instrument forward-folding fit and joint fit using flux-points employing a phenomenological model with two distinct emission components and three absorption components. \textbf{Top panel}: individual datasets fitted with simple models (power-laws, log-parabolas, power-laws with exponential cutoffs) and corresponding reconstructed flux points, and joint fits (flux points, forward folding). \textbf{Middle panel}: Residuals, calculated as the ratio of reconstructed differential spectrum to model prediction for the flux points fit (Residuals PF, open markers), and as the significance or ratio of the difference between the excess counts and model predicted counts to the square root of the model predictions for each bin in the forward folding case (Residuals FF, small filled circles). For clarity, and due to the large amount of energy bins in the X-ray datasets, only one every 5 bins is shown for NuSTAR, and one every 3 bins is shown for XRT. \textbf{Bottom panel}: Multiplicative absorption components.
    }
    \label{fig:Joint-fits}
\end{figure}

\begin{figure}
    \centering
    \includegraphics[width=1\linewidth]{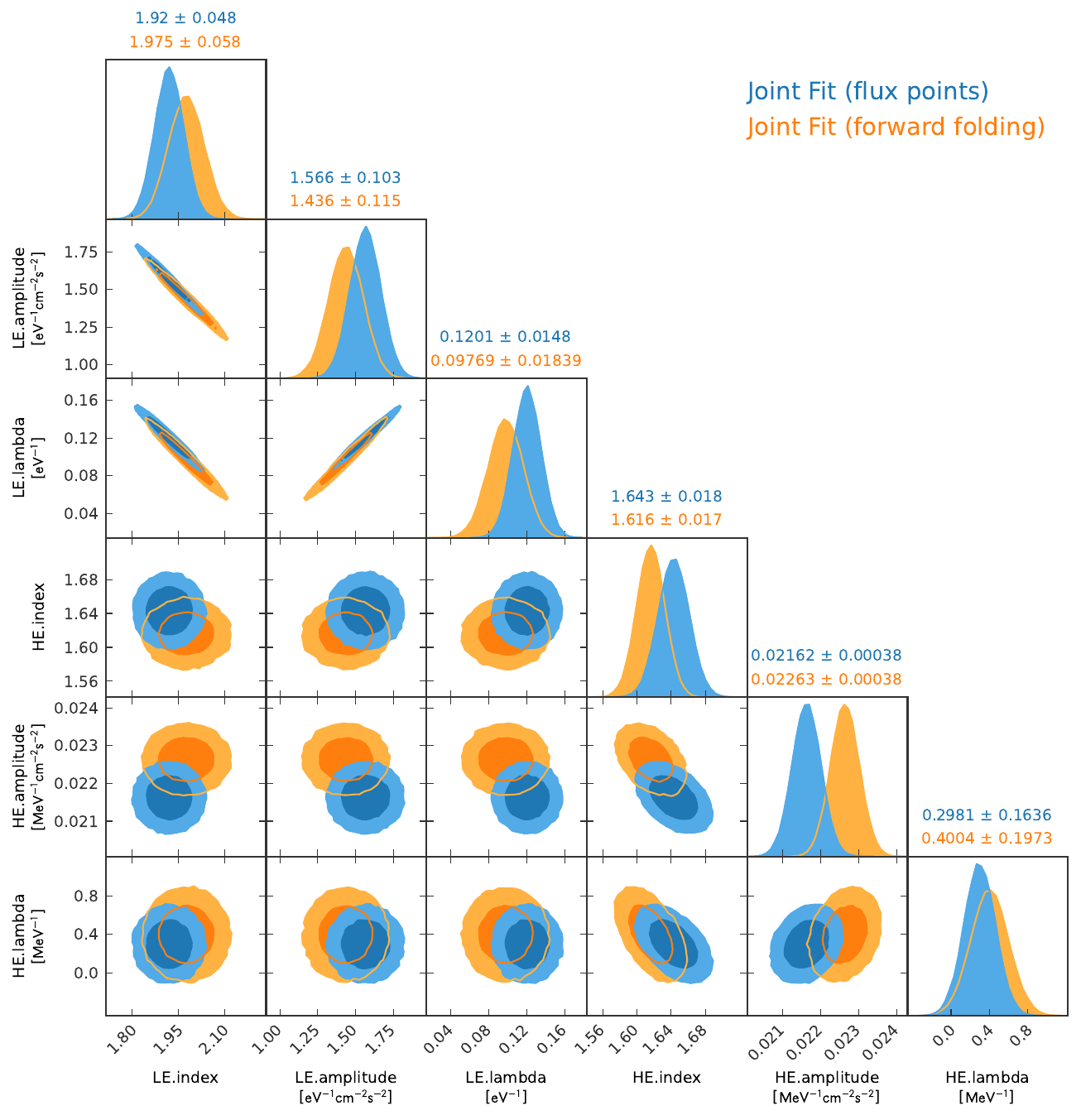}
    \caption{Corner plot showing the error ellipses calculated with a bootstrap method for the best-fit (free) parameters of the flux-point based fit and the full forward-folding joint fit. The model used for the LE is highly degenerate because of the limited amount of available data to constrain that component.} 
    \label{fig:Joint-fit-pars}
\end{figure}

Figure \ref{fig:Joint-fits} illustrates the best-fit models and their confidence bands for both fitting methods, while Figure \ref{fig:Joint-fit-pars} shows the correlations between the model parameters estimated from the covariance matrix for each fit. 
Both methods yield consistent results with well behaved residuals across the electromagnetic spectrum, confirming that the phenomenological model effectively captures the emission from OP~313 from optical to the $\gamma$-ray band. However, the forward folding technique shows better exploitation of the data, as it is able to use data beyond the last significant bins in both NuSTAR and {\it Fermi}-LAT. 

Other notable differences between the two methods include the treatment of instrumental effects and background components, more accurate in the forward folding case as the method directly incorporates energy redistribution and provides a more uniform representation of the emission in the observed counts space. In the NuSTAR case, bins for which the effective area of the instrument is approximately zero can still provide useful constraints to the instrumental background model. As opposed to directly fitting instrumental background components during the forward folding, the classical method inadequately performs a fit to `flux points' implying that the background is either fixed or subtracted. A similar problem occurs with {\it Fermi}-LAT skymodel, consisting on the sum of diffuse components (isotropic component, galactic diffuse emission) and a number of point-like source lying close to OP~313. In the classical method, the contributions of such components are `frozen', as opposed to fitted in the forward folding. Finally, the proposed workflow reduces systematic uncertainties associated with the conversion of counts to fluxes in the optical band as information coming from different filters is indirectly accounted for in the counts-flux conversion, likely improving the accuracy of the fit.

\subsection{Physical interpretation of the broadband emission}\label{sec:physical_interpretation}

The continuous spectral connection from X-rays to $\gamma$-rays observed in Figure \ref{fig:Joint-fits} suggests a coherent emission scenario likely originating from a single emission zone, and points to a single inverse Compton process to explain  both the X-ray and $\gamma$-ray emissions. 

The data also favors an external inverse Compton model for OP~313 due to the large ratio of peak energy fluxes in the high energy component of the SED to that of the low energy component (Compton Dominance). In a synchrotron-self-Compton scenario, achieving such high Compton Dominance would require too low magnetic fields and unrealistically high Doppler factors and/or electron densities. In contrast, the external Compton scenario can naturally result in large values for that ratio as electrons scatter the highly dense photon field from external components such as the accretion disk, the dusty torus or the broad line region of the AGN. 

To further identify the specific external radiation field responsible for the Compton up-scattering, additional data in the VHE regime would be necessary. Observations with instruments such as LST-1 or MAGIC could provide this information. This aspect is beyond the scope of this work, and will be explored in a forthcoming publication including VHE observations from these two instruments.

\section{Discussion}\label{sec:discussion}

In the preceding sections, we evaluated a flexible data format designed to streamline both archival and analysis processes. This format, based on OGIP and the GADF initiative, integrates direct observations — such as event counts, spectra (energy distributions of counts), or spatially-resolved data cubes — with IRFs from a diverse range of instruments, from space-borne $\gamma$-ray observatories like {\it Fermi}-LAT to ground-based optical telescopes like the LT.

\subsection{Flexibility and modeling capabilities}\label{sec:discussion:flexibility}

The method excels at providing flexibility for data modeling. We tested a phenomenological model with two simple 
power-laws with exponential cut-offs 
as emission components, aimed at representing synchrotron and inverse Compton emissions from a relativistic particle population in a jet. This model was complemented with fixed absorption components (extinction, neutral hydrogen, and EBL) and various background sources (diffuse radiation fields, additional sources in {\it Fermi}-LAT, and instrumental backgrounds in NuSTAR). 
Despite leaving free many of these additional components (normalizations of the spectral components of bright {\it Fermi}-LAT sources, {\it Fermi}-LAT isotropic and galactic diffuse components, and instrumental backgrounds in NuSTAR-A and NuSTAR-B),
the minimization converged within a reasonable computing time using the proposed forward folding technique, offering advantages over traditional methods that rely on reconstructed flux points.

The approach proposed here for multi-wavelength fitting is particularly beneficial for the X-ray data analysis part in this broadband context. 
Classical tools like {\tt xspec} are unfortunately often misused in the $\gamma$-ray community to produce flux points that are later fitted in the multi-wavelength context \citep[e.g.][]{agnpy,2022ApJ...934..158A} despite not being technically designed for that purpose. In the case of NuSTAR, this is especially relevant since its instrumental background model is a complex sum of multiple components, making it almost impossible to extract for the source of interest flux points using {\tt xspec}, as it would require to remove the contribution to the total flux from each of the other components. The proposed flexible framework, based on \texttt{gammapy}, removes the need of multi-wavelength flux points, and potentially enables a 3D analysis of XRT and NuSTAR data. The latter could be particularly advantageous for extended sources, or for a point source on top of an extended source (e.g. a halo), especially for instruments with limited angular resolution like NuSTAR. Therefore, the method tested in this work presents natural extensions and improvements.

\subsection{Main challenges and practical limits}

The increased flexibility and accuracy naturally comes with trade-offs. The method involves a more complex workflow and a larger parameter space. It also requires a more detailed modeling of observational data to incorporate all effects (astrophysical and instrumental backgrounds) and has to fold spectral models with various IRFs in each iteration to predict `observables' (counts, spectra, or data cubes) that are fit to the data. This complexity results in a higher computation time, and practical limits for the method’s application are yet to be fully assessed.

For {\it Fermi}-LAT, our analysis assumes that the PSF depends only on energy, and is constant within the FoV. This is a potential limitation with respect to its native analysis pipeline, {\it Fermitools}. Fortunately, for bright point-like sources like OP~313, where the contribution of other sources in the FoV is almost negligible, this effect should be small. Being \texttt{gammapy} an active developed tool, this drawback may be addressed in the future.

As photon-counting devices, {\it Swift}-XRT and {\it Swift}-UVOT are subject to the effects of pile-up and coincidence loss, which happen when the photon flux is large enough that two or more photons have a significant probability of being detected at the same position within a single CCD readout frame. In such cases, the detector registers the multiple events as a single event, with the recorded energy being the sum of the overlapping photons. This phenomenon skews the observed energy distribution and modifies the statistical properties of the data, deviating from pure Poissonian behavior and affecting the accuracy of count rate uncertainties. 

The treatment of these effects varies per instrument. For UVOT, a correction for coincidence loss is implemented in {\tt uvotsource}, effective as long as the field is not crowded and the source remains point-like \citep{2010MNRAS.406.1687B}. In the case of XRT, pile-up is less common in Windowed Timing (WT) mode unless the source is very bright, though it requires specialized treatment when it does occur \citep{2006NCimB.121.1521M}. In Photon Counting (PC) mode \citep{2006ApJ...638..920V}, a common approach to mitigate pile-up involves defining an annular extraction region to mask the brightest central pixels and estimate the source counts from the wings of the PSF. Like in UVOT, this approach is feasible only for isolated, point-like sources with minimal background contamination. More complex scenarios, such as extended sources or crowded fields, would likely require a more sophisticated spatial and spectral modeling of the emission and pile-up effects, well beyond the scope of this work.

\subsection{Considerations for optical instruments}

In the optical band, we tested a simplified IRF implementation tailored for relatively broad `filter' photometry. This approach represents a significant improvement over classical methods, which often just multiply the measured flux densities obtained through differential photometry techniques by an effective frequency of the band calculated under some assumption (typically flat photon spectrum or flat energy spectrum), ignoring the actual spectrum of the source. For steep spectra like that of OP~313 in the UV band, the real photon distribution is typically skewed, shifting the effective frequency of the band. 

The approximate IRFs we introduced in this work omit critical factors such as how atmospheric extinction, mirror reflectivity, aging of mirror coatings, and CCD defects modify the effective band profile. Including these elements in the IRFs --- though technically challenging --- would improve the accuracy of the spectral reconstruction. 
We did not explore spectral data analysis in the optical band due to the additional complexities involved, such as calibrating both wavelengths and fluxes, which could compromise the Poisson statistics of the observed data.

\section{Conclusions}\label{sec:conclusions}

The results presented in Section \ref{sec:results} demonstrate the effectiveness of the forward-folding technique for astrophysical data analysis in the context of complex multi-wavelength studies involving several instruments. This method not only shows that it is possible to organize, store, and analyze data from different instruments together in a consistent manner, but offers practical advantages over the classical flux-point fitting technique in terms of robustness. Furthermore, it allows for a more accurate statistical comparison of different modeling hypotheses, thereby simplifying the interpretation of the physical processes at play for a source like the FSRQ OP~313.

The resulting multi-instrument datasets and analysis code are stored in a format that is convenient for distribution and archiving. This format not only facilitates the reproduction of the results obtained in this specific study, but also supports further research on OP~313 with public datasets, as well as serve to make future validation tests for \texttt{gammapy}. Future work could involve detailed modeling of the spectral energy distribution using physical models through codes such as \texttt{jetset} \citep{jetset1,jetset2}, \texttt{agnpy} \citep{agnpy}, \texttt{Bjet\_MCMC} \citep{2024ApJ...962..140H,olivier_hervet_2024_13129296}, or \texttt{SOPRANO} \citep{2022MNRAS.509.2102G,2023mgm..conf..429G}, or exploring nested sampling methods \citep{2021JOSS....6.3001B}. 

This study provides a largely static view of OP~313 for two specific nights. Including variability studies and time-evolving models that account for particle injection and cooling will enhance our understanding of the emission of the source in particular and blazars in general. In this regard, the proposed simplified handling of instrument datasets will ensure greater consistency and reduce errors compared to classical methods that use data in disparate formats, as well as hopefully providing one of the first comprehensive multi-wavelength blazar datasets built over standardized data formats and open-source analysis tools.

\begin{acknowledgements}
We acknowledge the {\it Fermi}-LAT, NuSTAR, {\it Swift}-XRT, and {\it Swift}-UVOT teams for providing open-access data and analysis tools, as well as their support during this target-of-opportunity observation campaign. We also thank the Liverpool Telescope team for their prompt response and access to their robotic observations through reactive time.

We are also grateful to the \texttt{gammapy} team for their extremely versatile and open-source analysis pipeline, which played a fundamental role in this multi-instrument, multi-wavelength project.

M.N.R. acknowledges support from the {\it Agencia Estatal de Investigación del Ministerio de Ciencia, Innovación y Universidades} (MCIU/AEI) under grant PARTICIPACIÓN DEL IAC EN EL EXPERIMENTO AMS and the European Regional Development Fund (ERDF) with reference PID2022-137810NB-C22 / DIO 10.13039/501100011033. J.O.S. and D.M. acknowledge financial support from the projetct ref. AST22\_00001\_9 with founding from the European Union - NextGenerationEU, the \textit{Ministerio de Ciencia, Innovación y Universidades, Plan de Recuperación, Transformación y Resiliencia}, the \textit{Consejería de Universidad, Investigación e Innovación} from the Junta de Andalucía and the \textit{Consejo Superior de Investigaciones Científicas}. J.O.S. also acknowledges financial support through the Severo Ochoa grant CEX2021-001131-S funded by MCIN/AEI/ 10.13039/501100011033 and through grants PID2019-107847RB-C44 and PID2022-139117NB-C44 and founding from INFN Cap. U.1.01.01.01.009.
F.A. acknowledge support from the European Union’s Horizon 2020 Programme under the AHEAD2020 project (grant agreement n. 871158) and support by CNES, focused on methodology for X-ray analysis.

\end{acknowledgements}

%
%

\bibliographystyle{aa}  
\bibliography{aa} 

\begin{appendix} 

\section{Software description and data repository}\label{appendix:software_description}

This work is based on the open-source analysis tool \texttt{gammapy} and open data formats initiatives (GADF, OGIP). Only two new code implementations were required to conduct the analysis as presented in this manuscript. The first is a reader for standard OGIP files generated by HEASARC's native tools like {\tt uvot2pha}, {\tt xrtproducts}, and {\tt nuproducts}. This reader was originally developed by \citet{Giunti_gammapyXray} for XMM, and it was extended to also support NuSTAR, {\it Swift}-XRT, and {\it Swift}-UVOT. The second implementation is a set of \textsc{Python} classes that perform an optical point-like photometric reduction of Liverpool Telescope IO:O data, producing OGIP-compliant files from optical images using a set of reference stars with known photometry from Sloan's SDSS DR18 and PanSTARR's PS1 catalogs.

Additionally, we provide Jupyter notebooks \citep{2007CSE.....9c..21P,2014AGUFM.H44D..07R} to illustrate the details of the construction of the \texttt{gammapy}-compliant datasets, as well as the analysis of each individual dataset and the joint analysis of the broadband datasets. These codes, along with the resulting \texttt{gammapy}-compliant data files, are available in a GitHub repository \citep{Nievas_gammapy_mwl_workflow} for easy access and reproducible results. The structure of this repository is as follows:

\begin{itemize}
\item Notebooks/DatasetGenerator
\item Notebooks/DatasetAnalysis
\item Helpers
\item Models
\item Figures
\end{itemize}

Within the {\tt Notebooks/DatasetGenerator/} directory (with one subdirectory for each night of observations), we stored notebooks (one per instrument) that compile all the steps to generate the \texttt{gammapy}-compliant 1D and 3D binned datasets, as described in section \ref{sec:datasets}. Similarly, {\tt Notebooks/DatasetAnalysis/} contains notebooks with the analysis of the datasets from individual instruments and the joint analysis of the multi-wavelength (MWL) datasets. The {\tt Helpers/} directory contains functions for generating \texttt{gammapy} native multiplicative models for dust extinction, neutral hydrogen, and EBL absorption. It also includes classes for fetching photometry data from the PS1 and SDSS DR18 catalogs, as well as various file handling and plotting functions used across the analysis notebooks. The {\tt Models/} directory contains the tabular absorption models for EBL \citep{dominguez2011,saldana-lopez2021}, the neutral hydrogen absorption exported from {\tt xspec}'s {\tt TBabs} model \citep{2000ApJ...542..914W,heasarc_models}, and the dust extinction extracted from {\tt xspec}'s {\tt redden} \citep{1989ApJ...345..245C,heasarc_models}. Finally, the {\tt Figures/} directory (one subdirectory per observing night) contains plots and figures presented in this work.

\section{Reproducibility and Analysis Tools}

To ensure reproducibility of the multi-wavelength analysis presented in this work, detailed information is provided on the software tools (including versions), calibration files, observation IDs used for each instrument. The data were processed using publicly available software and calibrated with the most up-to-date instrument response files. The following tables summarize the relevant details for each instrument.

\subsection{Fermi-LAT}

Fermi-LAT data were analyzed using the \texttt{Fermitools} package version 2.2.0, using the latest available diffuse models for PASS8 and the most recent 4FGL-DR4 catalog release at the time. The analysis settings are shown in table \ref{tab:lat_analysis}.

\begin{table}
\centering
\caption{Details for Fermi-LAT data analysis.}
\begin{tabular}{p{3.2cm}p{4.8cm}}
\textbf{Parameter} & \textbf{Details} \\
\hline
Software & Fermitools 2.2.0 (Conda), enrico (git, April 2024) \\
\hline
Data release & PASS-8\\
\hline
Diffuse Model & gll\_iem\_v07, iso\_P8R3\_SOURCE\_V3\_v1\\
\hline
Catalog (skymodel) & 4FGL-DR4 (gll\_psc\_v35) \\
\hline
Sources (skymodel) & Those within $<25^\circ$ with Test Statistics $TS>1$, free spectral parameters if $<5^\circ$ \\
\end{tabular}
\label{tab:lat_analysis}
\end{table}

\subsection{NuSTAR}

NuSTAR observations were analyzed using the \texttt{nustardas} package version 2.1.2, as part of the {\tt HEASoft} 6.31.1 software suite. The instrumental background modeling was performed with \texttt{nuskybgd}. The observation IDs and analysis details are listed in table \ref{tab:nustar_analysis}

\begin{table}
\centering
\caption{Details for NuSTAR data analysis.}
\begin{tabular}{p{3.2cm}p{4.8cm}}
\textbf{Parameter} & \textbf{Details} \\
\hline
Software & HEASoft\,V6.31.1,\,nustardas\,2.1.2, nuproducts\,0.3.3 \\
\hline
Background model & nuskybgd 0.1.dev206 \\
\hline
CALDB & 20240729 \\
\hline
Obs. IDs & 91002609002\,(MJD\,60373), 91002609004\,(MJD\,60384) \\
\end{tabular}
\label{tab:nustar_analysis}
\end{table}

\subsection{Swift-XRT}

The X-ray Telescope (XRT) data were analyzed using the tools provided in the {\tt HEASoft} package, version 6.32.1. Specific details on the software versions, observations, and response files are reported in table \ref{tab:xrt_analysis}.

\begin{table}
\centering
\caption{Details for Swift-XRT data analysis.}
\begin{tabular}{p{3.2cm}p{4.8cm}}
\textbf{Parameter} & \textbf{Details} \\
\hline
Software & HEASoft\,6.32.1,\,xrtproducts\,0.4.3, XSPEC\,12.13.1 \\
\hline
CALDB & 20240729 \\
\hline
Response Matrices & swxpc0to12s6\_20210101v016 \\
\hline
Obs. IDs (OBSID) & 00036384074\,(MJD\,60373), 00089816002\,(MJD\,60384) \\
\end{tabular}
\label{tab:xrt_analysis}
\end{table}

\subsection{Swift-UVOT}

The UVOT data were processed using the UVOT software suite within {\tt HEASoft} 6.32.1, and the final data products were extracted using \texttt{uvot2pha}, which internally uses \texttt{uvotproduct} and \texttt{uvotsource}. Specific details are summarized in Table \ref{tab:uvot_analysis}.

\begin{table}
\centering
\caption{Details for Swift-UVOT data analysis.}
\begin{tabular}{p{3.2cm}p{4.8cm}}
\textbf{Parameter} & \textbf{Details} \\
\hline
Software & HEASoft\,6.32.1,\,uvotproduct\,2.9, uvotsource\,4.5,\,uvot2pha\,1.1 \\
\hline
CALDB & 20240729 \\
\hline
Response File (RSP) & 20041120v105 \\
\hline
Obs. IDs & 00036384074\,(MJD\,60373), 00089816002\,(MJD\,60384) \\
\end{tabular}
\label{tab:uvot_analysis}
\end{table}

\subsection{Liverpool IO:O}

The Liverpool IO:O was provided in already reduced format by Liverpool's automatic reduction, and the final photometric analysis was performed with a custom-made code, provided in the Notebooks folder of the GitHub repository, which is based on common python packages like {\tt astropy} or {\tt photutils}, and public catalogs from Sloan and Pan-STARRS

\begin{table}[h]
\centering
\caption{Details for Liverpool IO:O data analysis.}
\begin{tabular}{p{3.2cm}p{4.8cm}}
\textbf{Parameter} & \textbf{Details} \\
\hline
Software & photutils\,1.8.0,\,astropy\,5.3.4, astroquery\,0.4.7, gammapy\,1.3.dev726+g6c6d15956 \\
\hline
Catalogs & Pan-STARRS\,PS1-DR2, SDSS\,DR18 \\
\end{tabular}
\end{table}

\FloatBarrier

\section{Dataset analysis validation}\label{appendix:validation}

This appendix shows the comparison between the analyses conducted using native tools specific to each instrument (i.e. {\tt xspec} for NuSTAR and XRT, {\tt Fermitools}/{\tt enrico} for {\it Fermi}-LAT) and those performed on the OGIP- and GADF-compliant datasets using \texttt{gammapy}. This comparison is needed for validating the consistency and reliability of the \texttt{gammapy}-based analysis, and will turn crucial to later extend the full statistical forward folding technique across different wavelength regimes.

\begin{figure*}
    \centering
    \begin{subfigure}[b]{0.315\linewidth}
        \centering
        \includegraphics[height=0.21\textheight,keepaspectratio]{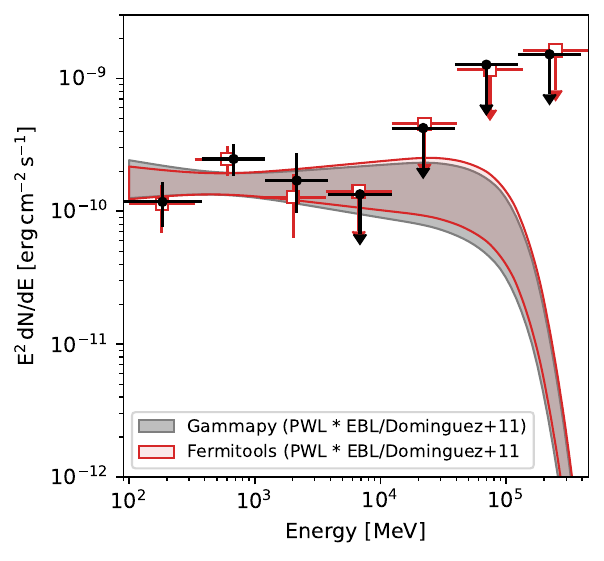}
        \caption{{\it Fermi}-LAT.}
        \label{fig:fermi_bestfit}
    \end{subfigure}
    \begin{subfigure}[b]{0.315\linewidth}
        \centering
        \includegraphics[height=0.217\textheight,keepaspectratio]{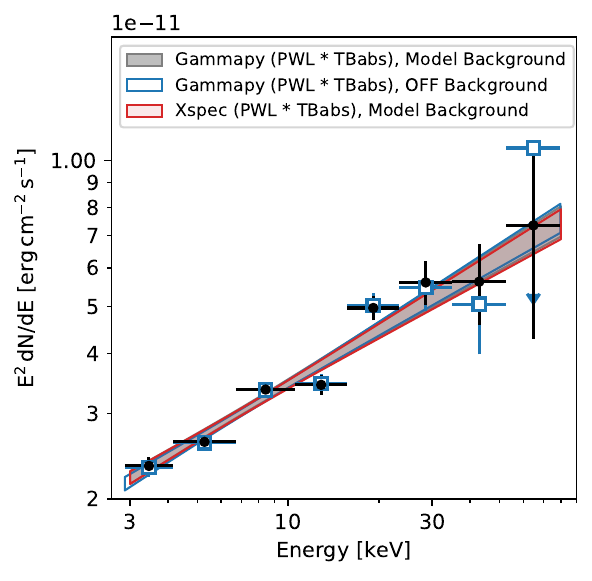}
        \caption{NuSTAR}
        \label{fig:nustar_bestfit}
    \end{subfigure}
    \begin{subfigure}[b]{0.315\linewidth}
        \centering
        \includegraphics[height=0.217\textheight,keepaspectratio]{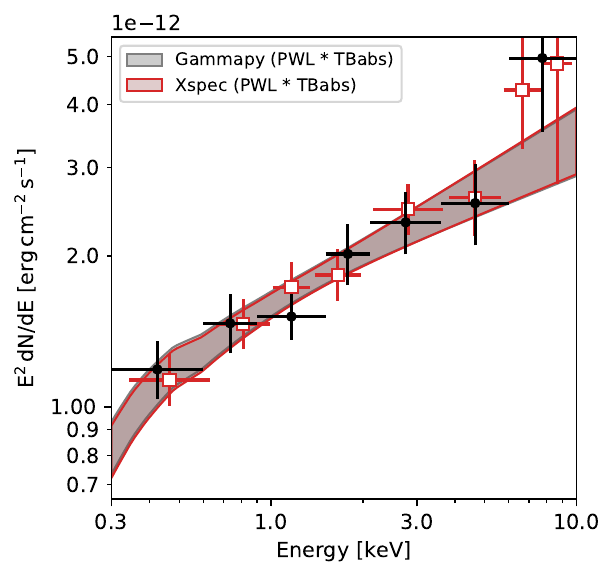}
        \caption{{\it Swift}-XRT}
        \label{fig:xrt_bestfit}
    \end{subfigure}
    \caption{Best-fit power-law models (with EBL absorption in {\it Fermi}-LAT and neutral hydrogen absorption for NuSTAR and {\it Swift}-XRT) fitted to each dataset with native tools (\texttt{xspec} or Fermitools depending on the case) and \texttt{gammapy}. For reference, the spectral reconstruction with the OFF Background estimate in NuSTAR (second panel) is shown in blue for reference. Upper limits are calculated at a 95\% confidence level from the {\tt cstat/wstat} likelihood profiles. } 
    \label{fig:best_fits}
\end{figure*}

For high-energy instruments like {\it Fermi}-LAT, NuSTAR, and \textit{Swift}-XRT, we directly compared in Figure \ref{fig:best_fits} the best-fit spectral energy distribution confidence bands and corresponding flux points. To keep the comparison straightforward and the parameter space limited to just two variables, we used a simple power-law in each band, with either EBL absorption in $\gamma$ rays or neutral hydrogen absorption in X-rays. Even though  {\it Fermi}-LAT $\gamma$-ray data hints significant curvature in that band, using a simple model is enough to test the validity of the method, and only a power-law with exponential cut-off was used in Figure \ref{fig:Joint-fits} for the broadband analysis. 

The results shown in figures \ref{fig:fermi_bestfit}, \ref{fig:nustar_bestfit} and \ref{fig:xrt_bestfit} indicate that the best-fit model and contours obtained with \texttt{gammapy} are indistinguishable from those obtained with the native tools, demonstrating the potential of \texttt{gammapy} as a multi-wavelength data analysis framework. 

\begin{figure*}
\centering
    \begin{subfigure}[b]{0.32\linewidth}
        \centering
        \includegraphics[height=0.22\textheight,keepaspectratio]{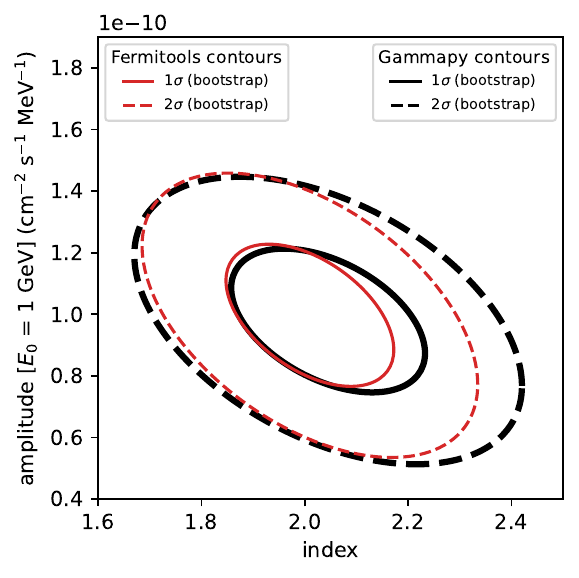}
        \caption{{\it Fermi}-LAT.}
        \label{fig:fermi_ellipses}
    \end{subfigure}
    \begin{subfigure}[b]{0.32\linewidth}
        \centering
        \includegraphics[height=0.22\textheight,keepaspectratio]{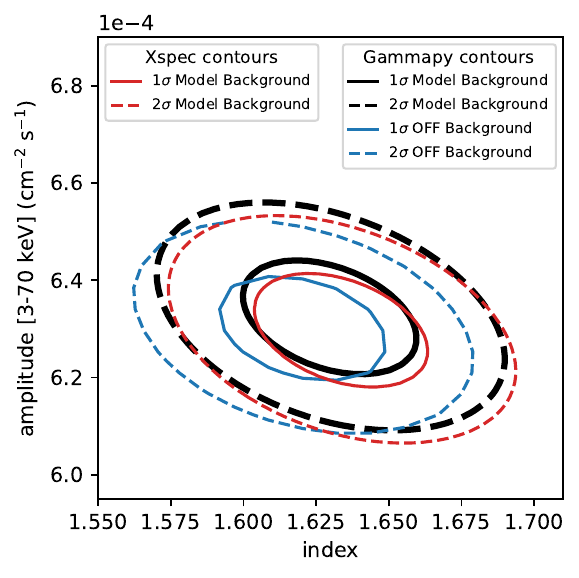}
        \caption{NuSTAR}
        \label{fig:nustar_ellipses}
    \end{subfigure}
    \begin{subfigure}[b]{0.32\linewidth}
        \centering
        \includegraphics[height=0.22\textheight,keepaspectratio]{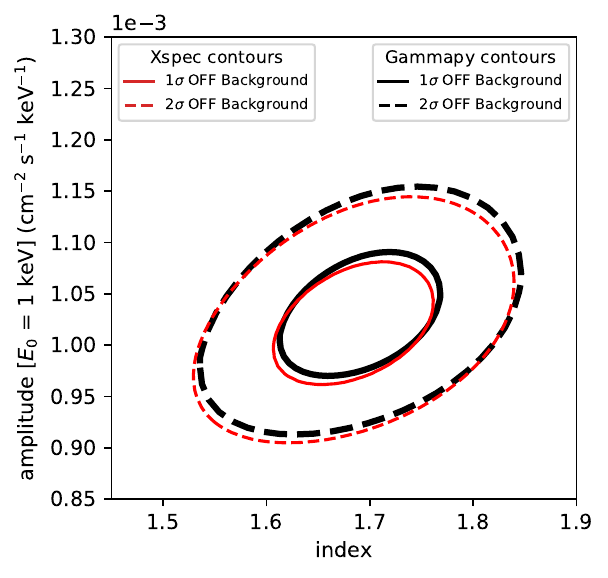}
        \caption{{\it Swift}-XRT}
        \label{fig:xrt_ellipses}
    \end{subfigure}
    \caption{Parameter error ellipses on the amplitude (integral flux for NuSTAR) and spectral index for the power-law fit to the {\it Fermi}-LAT, NuSTAR A+B and {\it Swift}-XRT datasets. Contours in black correspond to \texttt{gammapy}, while the native tool contours are shown in red. For NuSTAR, the contours obtained with the Poisson or OFF background estimate are shown in blue for reference.}
    \label{fig:error_ellipses}
\end{figure*}

In addition, we show in Figure \ref{fig:nustar_bestfit} the spectral reconstruction with both the instrumental background and the Poisson background estimate, matching well below $\sim 30$\, keV. We note however that {\tt xspec} has two severe limitations. The first is linked to the definition of a the power-law, whose pivot energy in {\tt xspec} is fixed to $1\,$keV and cannot be changed. To avoid impacting the results with a reference energy that is outside the energy range of NuSTAR, we redefined the power-law so that the normalization is done with respect to the the integral flux in the $3-70\,$keV band, instead of at a specific energy. The second limitation is that {\tt xspec} is not designed to produce flux points as we noted in Section \ref{sec:discussion:flexibility}, and only in simple cases with few spectral components (NuSTAR and {\it Swift}-XRT when doing an OFF background analysis with Poisson background statistics) it was feasible to estimate some sort of flux points for different energy bins. For more sophisticated cases, such as NuSTAR with the complex instrumental background model produced by {\tt nuskybgd}, the flux points estimated with {\tt xspec} do not correspond to the source component but to the total sky model, including the systematic background. Removing the additional background components (which are not fixed) is not trivial, therefore we only report in Figure \ref{fig:nustar_bestfit} the SED confidence band from {\tt xspec}, without the flux points. We also checked the parameter error contours in the amplitude vs spectral index parameter space in Figure \ref{fig:error_ellipses}, noting a very good agreement for the three instruments.

\begin{table*}
    \caption{Comparison of parameter values for the best-fits of the native tools (fermitools+minuit/migrad for {\it Fermi}-LAT data, \texttt{xspec}'s Levenberg-Marquardt algorithm for NuSTAR and {\it Swift}-XRT) with the same results obtained with \texttt{gammapy} with the default minuit/migrad as minimizer and algorithm. For simplicity, a classical power-law with absorption with amplitude defined as the photon density at a reference energy was used for \textit{Fermi}-LAT and \textit{Swift}-XRT (including also EBL absorption for \textit{Fermi}-LAT following in this case \cite{dominguez2011} as it is included in the {\tt Fermitools}, and neutral hydrogen for \textit{Swift}-XRT as implemented in {\tt xspec}'s {\tt TBabs} code). For NuSTAR, a power-law with integral as amplitude parameter was utilised as \texttt{xspec} does not allow to change the default reference energy, which is fixed at $1\,$keV and therefore out of the energy range of the instrument. Switching to migrad in {\tt xspec} showed a completely negligible impact on the estimated minima.}
    \label{tab:parameters}
    \centering
    \begin{tabular}{lcccccc}
         & \multicolumn{2}{c}{\textbf{\textit{Fermi}-LAT} (at $1\,$GeV)} & \multicolumn{2}{c}{\textbf{NuSTAR A+B} ($3-70\,$ keV)} & \multicolumn{2}{c}{\textbf{\textit{Swift}-XRT} (at $1\,$keV)} \\
        \cline{2-7}
         & amplitude & index & integral flux & index & amplitude & index \\
        \hline
        native  & $1.00 \pm 0.23$ & $2.01\pm 0.16$ & $6.30\pm 0.12$ & $1.633\pm0.030$ & $10.20\pm0.60$ & $1.683 \pm 0.077$  \\
        \texttt{gammapy} & $0.98 \pm 0.23$ & $2.05\pm 0.19$ & $6.32\pm 0.12$ & $1.630 \pm 0.030$ & $10.29\pm0.60$ & $1.690 \pm 0.078$  \\
        \hline
        scale/units & $\times 10^{-11}$ & $\mathrm{MeV^{-1} s^{-1} cm^{-2}}$ & $\times 10^{-4}$ & $\mathrm{s^{-1} cm^{-2}}$ & $\times 10^{-4}$ & $\mathrm{keV^{-1} s^{-1} cm^{-2}}$ \\
    \end{tabular}
\end{table*}

\begin{figure}
\centering    
    \begin{subfigure}[b]{0.495\linewidth}
        \centering
        \includegraphics[height=0.182\textheight,keepaspectratio]{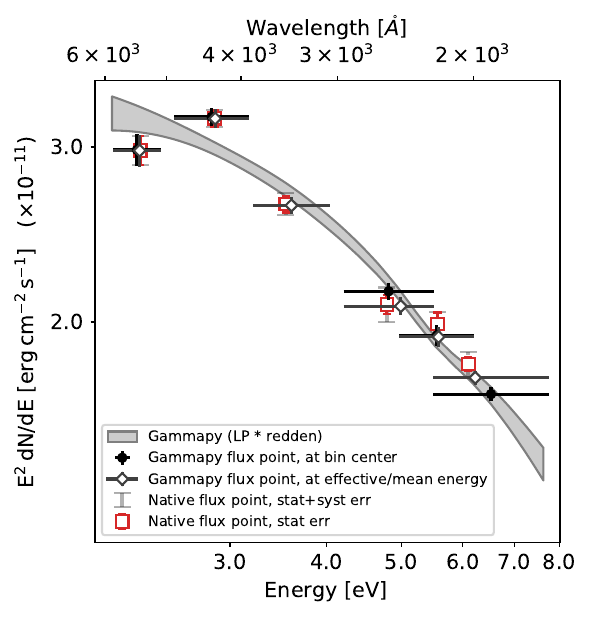}
        \caption{{\it Swift}-UVOT}
        \label{fig:uvot_bestfit}
    \end{subfigure}
    \begin{subfigure}[b]{0.495\linewidth}
        \centering
        \includegraphics[height=0.182\textheight,keepaspectratio]{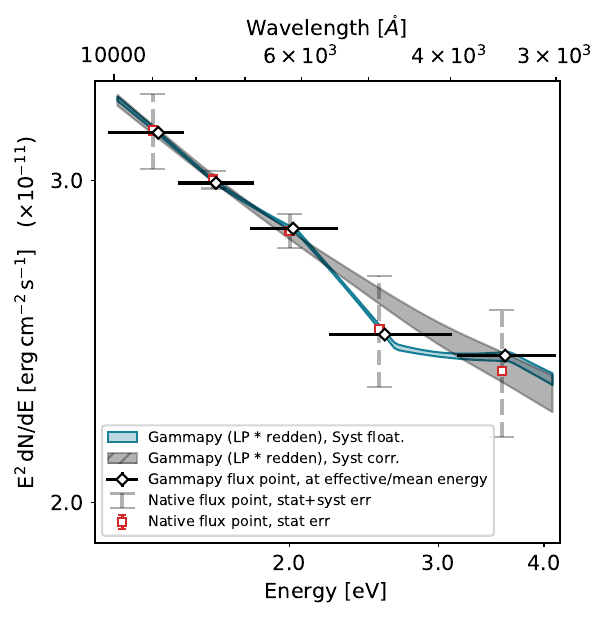}
        \caption{Liverpool Telescope IO:O}
        \label{fig:liverpool_bestfit}
    \end{subfigure}
    \caption{Best-fit spectra (log-parabola with dust extinction) and flux points reconstructed using \texttt{gammapy} --- based on counts and IRFs --- and with native tools using flux densities multiplied by the effective filter frequency center. Flux points in \texttt{gammapy} are evaluated over the geometrical center of the bin (black filled circles) and at the effective mean energy of the band with open diamond symbols (under the assumption of a flat source photon spectrum, to be roughly comparable to the classical method). LT datasets from the same filter were stacked together using \texttt{gammapy}'s {\tt stack\_reduce} method, that co-adds the counts, background and exposures for datasets with compatible energy axes. For the native analysis, weighted averages are shown instead as red boxes. Dashed gray error bars depict the total error once incorporating systematic uncertainties.}
    \label{fig:optical_sedpoints}
\end{figure}

\begin{figure}
\centering    
    \begin{subfigure}[b]{0.495\linewidth}
        \centering
        \includegraphics[height=0.163\textheight,keepaspectratio]{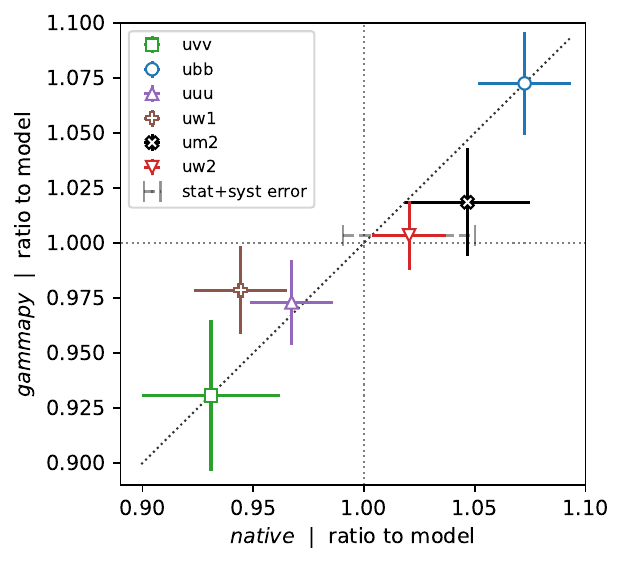}
        \caption{{\it Swift}-UVOT}
        \label{fig:uvot_sedpoints_comp}
    \end{subfigure}
    \begin{subfigure}[b]{0.495\linewidth}
        \centering
        \includegraphics[height=0.163\textheight,keepaspectratio]{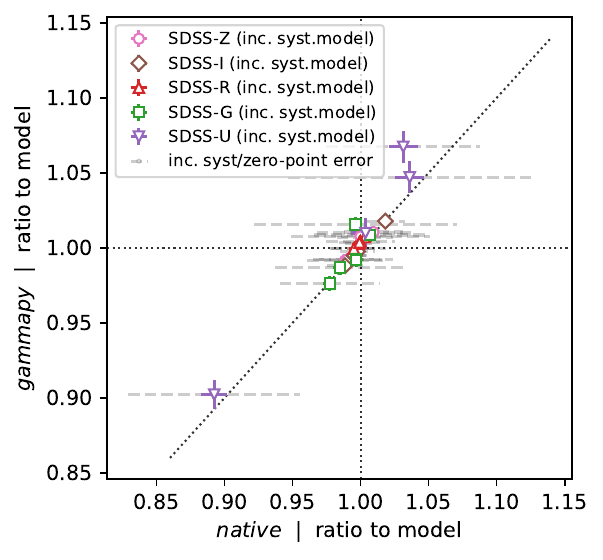}
        \caption{Liverpool Telescope IO:O}
        \label{fig:liverpool_sedpoints_comp}
    \end{subfigure}
    \caption{Comparison of reconstructed `relative' flux values using native photometry tools (reconstructed flux densities multiplied by the filter's effective frequency) and those obtained with \texttt{gammapy} (using counts and instrument response functions). All `relative' fluxes are shown as a ratio to the best-fit \texttt{gammapy} model of the joint datasets, evaluated at the corresponding point energy.}
    \label{fig:optical_sedpoints_comp}
\end{figure}

Finally, to validate the analysis on single-channel datasets, we estimated the all-channel best-fit spectral model for both {\it Swift}-UVOT and {\it Liverpool} IO:O, together with reconstructed flux points for each filter individually. The latter are compared to the flux derived from the native analysis, i.e. the output of \texttt{uvotsource} for {\it Swift}-UVOT, and the differential photometry fluxes reconstructed with our \texttt{photutils}-based pipeline for {\it Liverpool} IO:O. The resulting SEDs are shown in figures \ref{fig:uvot_bestfit} and \ref{fig:liverpool_bestfit} respectively. To reduce the effect introduced by the ambiguity in the selection of the reference energy of the band (estimated differently for \texttt{gammapy} and the native analyses), figures \ref{fig:uvot_sedpoints_comp} and \ref{fig:liverpool_sedpoints_comp} compare the ratio of each flux point over the best model prediction at the reference energy of the flux point.

\section{Background estimation for NuSTAR}
\label{appendix:background_nustar}

Two different methods for estimating the background in 1D datasets were explored for NuSTAR observations. The first method is the general Poisson or OFF background estimate.
This method makes minimal assumptions about the structure of the background spectra, primarily that the background spectrum is identical in the OFF region as in the ON region. While this approach offers significant flexibility, it has the downside that Poisson fluctuations can dominate the often weak emission from the source, particularly at higher energies. Moreover, the assumption of Poisson statistics requires all background spectral bins to have non-zero counts to ensure a correctly defined likelihood function. This can pose a problem for photon-counting instruments like NuSTAR, especially when integrating over short exposure times. To mitigate issues arising from zero-count bins, it often becomes necessary to group the data into coarser energy bins, which can compromise spectral resolution.

The second method involves modeling the background, a technique extensively used in the analysis of data from instruments like {\it Fermi}-LAT, where detailed background models such as diffuse galactic emission maps and isotropic emission spectra have been developed. In this work, we employ a similar approach for NuSTAR by using instrumental background spectral models generated by the \texttt{nuskybgd} tool. This method 
provides an alternative to the Poisson background estimate by incorporating a more structured model of the instrumental background into the analysis.

A direct comparison between the reconstructed spectra of the FSRQ OP~313 obtained using the two methods is shown in Figure \ref{fig:nustar_bestfit}, with the OFF background estimation results depicted in blue and the model background prediction results in black. Additionally, Figure \ref{fig:nustar_ellipses} presents the best-fit parameter error ellipses for both methods. The comparison reveals that at lower energies, where photon statistics are robust, the results from both methods are fully compatible. However, discrepancies begin to emerge at energies above 30 keV. Notably, in the 60–70 keV range, the OFF background estimate only provides an upper limit (with less than a 2$\,\sigma$ excess), while the model background based analysis continues to reconstruct a flux point with a significance exceeding 2$\,\sigma$.

These findings suggest that while both methods are viable, the choice of background estimation method can significantly impact the results at higher energies, particularly when the source signal is weak compared to the background. The model background method, with its more sophisticated modeling of the instrumental and astrophysical background, may offer advantages in these regimes, potentially yielding more accurate spectral reconstructions where the OFF background method falls short.

\FloatBarrier

\section{Systematic Uncertainties in Liverpool Telescope Data}
\label{appendix:optical_photometry_systematics}

In the classical analysis, systematic uncertainties related to errors in the zero-point estimation (star-to-star variations) are typically propagated in quadrature to the final error estimate of the source flux. While this method is model-independent, it does not take full advantage of the detailed broadband spectrum of the source, which becomes more apparent as additional data across different energy bands are incorporated.

In this work, we explore an alternative approach using the {\tt PiecewiseNormSpectralModel} in \texttt{gammapy}. This model adds an energy-dependent multiplicative component to our source's spectral model, allowing the normalization parameters at each fixed energy node to be free. To avoid creating a highly degenerate model by leaving these parameters completely unconstrained, we impose Gaussian priors on each normalization factor. These priors are centered at 1, with widths corresponding to the systematic uncertainties arising from star-to-star variations in the calculated exposure.

When this component together with the source spectrum is fitted to the data, it causes a deformation of the assumed spectrum (e.g., a log-parabola), enabling the model to better align with the actual measurements (excess counts) by subtly adjusting the shape of the spectrum (see Figure \ref{fig:liverpool_bestfit}, light gray curve and black diamond markers).
By removing this component, we can recover the underlying estimate of the true spectrum (dark gray), free from the influence of systematic errors. For clarity, in this figure we stacked together observations using the same filter, which in \texttt{gammapy} it is technically implemented in the {\tt stack\_reduce} method by co-adding the counts, background, and exposures for datasets with compatible axes. The equivalent stacking is achieved for the native (classical) analysis by performing a weighted average of the flux points, where the weights are set to the inverse of the square of the statistical uncertainties of the points. 

\FloatBarrier

\section{Second night analysis (MJD60384)}\label{appendix:second_night}

\begin{figure}
    \centering
    \includegraphics[width=1\linewidth]{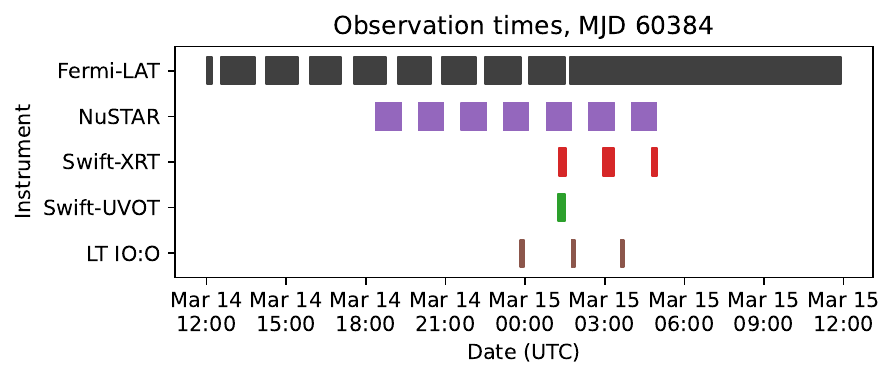}
    \caption{Temporal coverage of the second observation night.}
    \label{fig:dataset_gti2}
\end{figure}

The body of this manuscript primarily discusses the methodology used to create \texttt{gammapy}-compatible datasets, with data from the first night of NuSTAR observations (March 4th, 2024, MJD 60373) presented as an example of the reconstruction achievable using our analysis framework. In this appendix, we briefly report the results of applying the same analysis to the second night of NuSTAR observations (March 15th, 2024, MJD 60384). The corresponding temporal coverage is still  consistent across all instruments, as seen in Figure \ref{fig:dataset_gti2}. 

\begin{figure}
    \centering
    \includegraphics[width=1\linewidth]{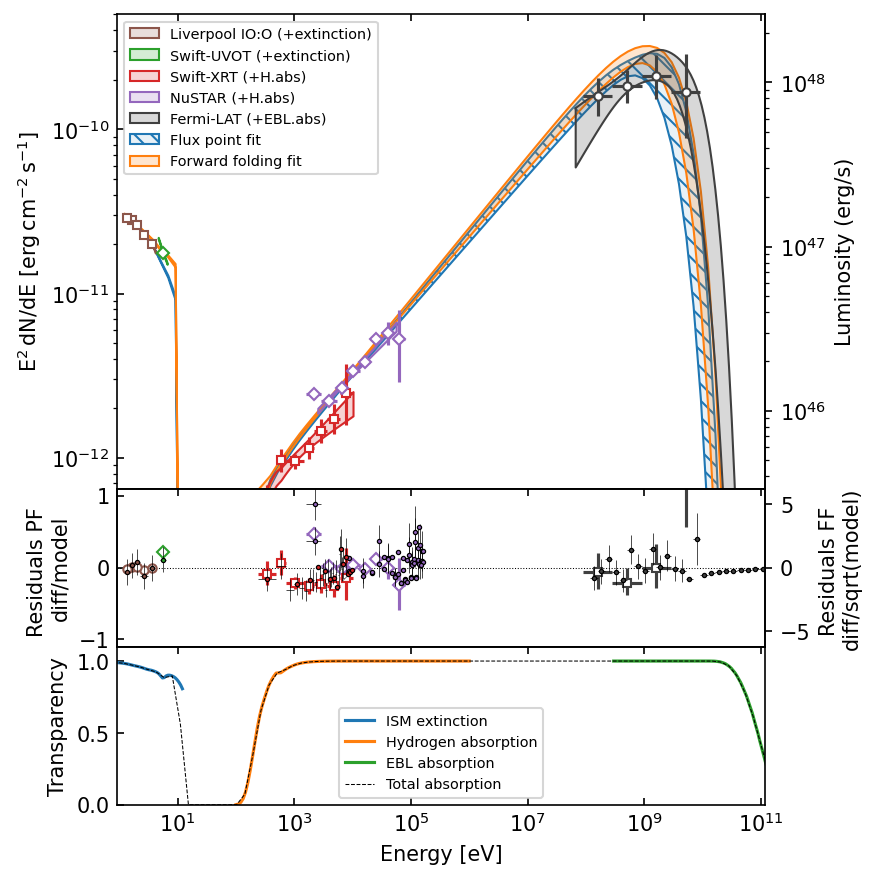}
    \caption{Joint multi-instrument forward-folding fit and joint fit using flux-points from the night of March 15, 2024 (MJD60384) with the similar structure as in Figure \ref{fig:Joint-fits}. Note that {\it Swift} data were affected by a severe ``loss of lock'' problem that night.}
    \label{fig:Joint-fits2}
\end{figure}

To keep this section short and focused on the final results, we focus on the two main figures from Section  \ref{sec:physical_interpretation}. The first figure is the best-fit models, shown in Figure \ref{fig:Joint-fits2}, using the same sum of two power-laws with exponential cut-offs as in Figure \ref{fig:Joint-fits}. 
We note that {\it Swift}-UVOT has a single observation with filter UM2, but more importantly, as discussed in section \ref{sec:datasets:uvot}, {\it Swift} was affected by a noticeable degradation in the pointing accuracy caused by noise in one gyroscope \citep{gcn1,gcn2}. This is in fact hinted in our {\it Swift}-XRT data, whose integrated PC-mode image shows a larger PSF than usual, and it is readily visible in UVOT, where OP~313 images are elongated and distorted, requiring an abnormally large source integration region of $25"$ radius to fully contain the signal. The standard $5"$ radius yields an underestimation of the source flux by a factor $\sim 2$. 

\begin{figure}
    \centering
    \includegraphics[width=1\linewidth]{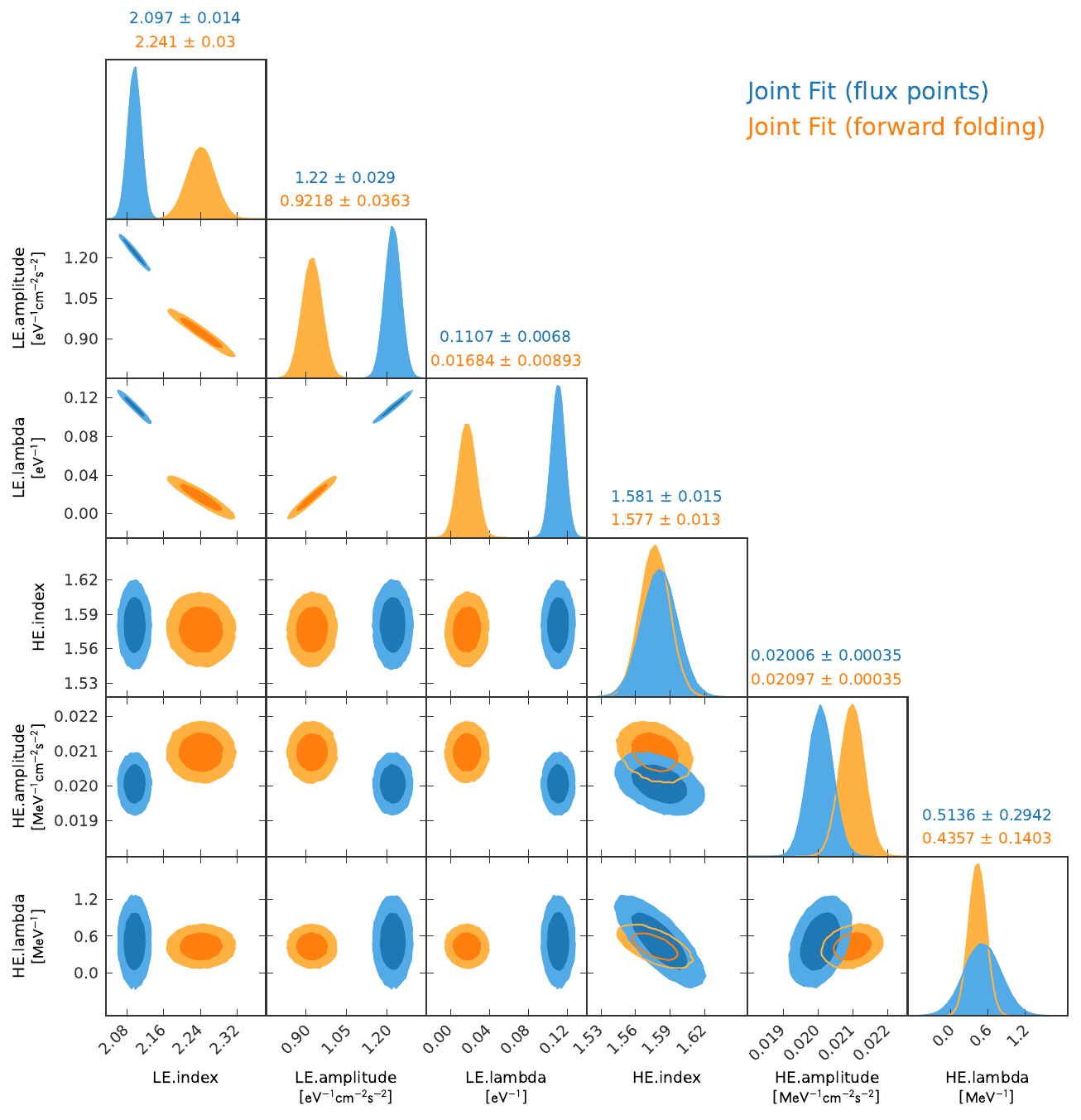}
    \caption{Corner plot showing the error ellipses calculated with a bootstrap method for the best-fit (free) parameters of the flux-point based fit and the full forward-folding joint fit. The model used for the LE is highly degenerate because of the limited amount of available data to constrain that component.} 
    \label{fig:Joint-fit-pars2}
\end{figure}

The corresponding corner plot, representing the error ellipses for each pair of free model parameters for the source emission component, is displayed in Figure \ref{fig:Joint-fit-pars2}. Compared to Figure \ref{fig:Joint-fit-pars}, we observe a similar degree of degeneracy between the parameters for the low-energy (LE) component. 
Moreover, despite our efforts to enlarge the source extraction region in both {\it Swift}-XRT and {\it Swift}-UVOT to accommodate their poorer PSF, a notable discrepancy persists between {\it Swift}-XRT and the spectral reconstruction from NuSTAR and {\it Fermi}-LAT. This inconsistency leads to less stable fitting process and different best-fit solutions for the low-energy component between the flux point fitting analysis and the forward folding technique.

It is worth noting that both the significant model degeneracies and fit convergence issues are common challenges faced when working with large and diverse datasets. However, these issues could potentially be mitigated by adopting a modern, fully Bayesian nested sampling inference engine such as {\tt UltraNest} \citep{2021JOSS....6.3001B}, which efficiently explores the entire prior parameter space to build robust posterior probability distributions for model parameters, even if the distribution is multi-modal. With the full dataset exported into a standard framework and the construction of a full forward folding likelihood, we are already halfway to exploring this approach.

\end{appendix}

\end{document}